\documentclass[fleqn,11pt,twoside]{article}
\usepackage{amsthm,amssymb, color, epsfig, graphics, subfigure}

\usepackage{tikz}

\usepackage{graphicx}
\usepackage{latexsym}
\usepackage[all]{xy}
\usepackage{amsmath, amscd}
\usepackage{graphicx}
\usepackage{lscape}

\newtheorem{proposition}{Proposition}

\makeatletter
\newcommand{\copyrightnote}[2]{{\renewcommand{\thefootnote}{}
 \footnotetext{\small\it
\begin{flushleft}
Copyright \copyright \ #1 by  #2
\end{flushleft}}}}

\newcommand{\Name}[1]{\begin{flushleft}
                       \LARGE \bf #1
                       \end{flushleft}\vspace{-3mm}}

\newcommand{\Author}[1]{\begin{flushleft}
                       \it #1 \end{flushleft}}

\newcommand{\Address}[1]{\begin{flushleft}
                       \it #1 \end{flushleft}}

\newcommand{\Date}[1]{\begin{flushleft}
                      \small  \it #1 \end{flushleft}}

\newcommand{\evenhead}{Author \ name}
\newcommand{\oddhead}{Article \ name}

\renewcommand{\@evenhead}{
\hspace*{-3pt}\raisebox{-15pt}[\headheight][0pt]{\vbox{\hbox to \textwidth
{\thepage \hfil \evenhead}\vskip4pt \hrule}}}
\renewcommand{\@oddhead}{
\hspace*{-3pt}\raisebox{-15pt}[\headheight][0pt]{\vbox{\hbox to \textwidth
{\oddhead \hfil \thepage}\vskip4pt\hrule}}}
\renewcommand{\@evenfoot}{}
\renewcommand{\@oddfoot}{}

\long\def\@makecaption#1#2{%
  \vskip\abovecaptionskip
  \sbox\@tempboxa{\small \textbf{#1.}\ \ #2}%
  \ifdim \wd\@tempboxa >\hsize
    {\small \textbf{#1.}\ \ #2}\par
  \else
    \global \@minipagefalse
    \hb@xt@\hsize{\hfil\box\@tempboxa\hfil}%
  \fi
  \vskip\belowcaptionskip}

\newcommand{\JNMPnumberwithin}[3][\arabic]{%
  \@ifundefined{c@#2}{\@nocounterr{#2}}{%
    \@ifundefined{c@#3}{\@nocnterr{#3}}{%
      \@addtoreset{#2}{#3}%
      \@xp\xdef\csname the#2\endcsname{%
        \@xp\@nx\csname the#3\endcsname .\@nx#1{#2}}}}%
}

\newcommand{\resetfootnoterule} {
  \renewcommand\footnoterule{%
  \kern-3\p@
  \hrule\@width.4\columnwidth
  \kern2.6\p@}
}

\renewcommand{\footnoterule}{}

\newcommand{\be}{\begin{equation}}
\newcommand{\ee}{\end{equation}}
\newcommand{\ba}{\hspace*{-5pt}\begin{array}}
\newcommand{\ea}{\end{array}}
\newcommand{\p}{\partial}

\makeatother
\numberwithin{equation}{section}
\theoremstyle{definition}

\theoremstyle{proposition}

\renewcommand{\ba}{\begin{array}}
\renewcommand{\ea}{\end{array}}
\newcommand{\beg}{\begin{eqnarray}}
\newcommand{\eeq}{\end{eqnarray}}
\newcommand{\bg}{\begin{eqnarray*}}

\newcommand{\ed}{\end{eqnarray*}}

\newcommand{\nn}{\nonumber}

\renewcommand{\p}{\partial} 
 
\newcommand{\notlhd}{\lhd\kern-.8em{/}\ } 
\newcommand{\notexist}{\ \exists\kern-.5em{\raise.1em\hbox{/}}\ }

\newcommand{\pde}[2]{\frac{\p #1}{\p #2}} 
 
\newcommand{\pdd}[2]{\frac{\p^2 #1}{\p #2^2}} 
\newcommand{\inp}{{\mbox{\vbox{\hrule width0ex\hbox{\vrule
 height0ex\kern3.8pt
\vbox{\kern2.5pt}\kern3.8pt \vrule height1.6ex}
\hrule width1.6ex}}}}

\makeatletter
\newcommand*{\da@rightarrow}{\mathchar"0\hexnumber@\symAMSa 4B }
\newcommand*{\da@leftarrow}{\mathchar"0\hexnumber@\symAMSa 4C }
\newcommand*{\xdashrightarrow}[2][]{%
  \mathrel{%
    \mathpalette{\da@xarrow{#1}{#2}{}\da@rightarrow{\,}{}}{}%
  }%
}
\newcommand{\xdashleftarrow}[2][]{%
  \mathrel{%
    \mathpalette{\da@xarrow{#1}{#2}\da@leftarrow{}{}{\,}}{}%
  }%
}
\newcommand*{\da@xarrow}[7]{%
  \sbox0{$\ifx#7\scriptstyle\scriptscriptstyle\else\scriptstyle\fi#5#1#6\m@th$}%
  \sbox2{$\ifx#7\scriptstyle\scriptscriptstyle\else\scriptstyle\fi#5#2#6\m@th$}%
  \sbox4{$#7\dabar@\m@th$}%
  \dimen@=\wd0 %
  \ifdim\wd2 >\dimen@
    \dimen@=\wd2 %   
  \fi
  \count@=2 %
  \def\da@bars{\dabar@\dabar@}%
  \@whiledim\count@\wd4<\dimen@\do{%
    \advance\count@\@ne
    \expandafter\def\expandafter\da@bars\expandafter{%
      \da@bars
      \dabar@ 
    }%
  }%  
  \mathrel{#3}%
  \mathrel{%   
    \mathop{\da@bars}\limits
    \ifx\\#1\\%
    \else
      _{\copy0}%
    \fi
    \ifx\\#2\\%
    \else
      ^{\copy2}%
    \fi
  }%   
  \mathrel{#4}%
}
\makeatother
\begin{document}

\renewcommand{\evenhead}{ {\LARGE\textcolor{blue!10!black!40!green}{{\sf \ \ \ ]ocnmp[}}}\strut\hfill 
M Euler, N Euler and F Oliveri
}
\renewcommand{\oddhead}{ {\LARGE\textcolor{blue!10!black!40!green}{{\sf ]ocnmp[}}}\ \ \ \ \  
Differential equations invariant under a projective transformation
}

%%%% Matter for the first page 
\thispagestyle{empty}
\newcommand{\FistPageHead}[3]{
\begin{flushleft}
\raisebox{8mm}[0pt][0pt]
{\footnotesize \sf
\parbox{150mm}{{\textcolor{blue!10!black!40!green}{{\bf Open Communications in Nonlinear Mathematical Physics}}}
\ \ {Special Issue: Bluman}, 2025\\[0.1cm]
\strut\hfill 
ocnmp:16901,
pp #2\hfill {\sc #3}}}\vspace{-13mm}
\end{flushleft}}

\FistPageHead{1}{\pageref{firstpage}--\pageref{lastpage}}{ \ \ }

\strut\hfill

\strut\hfill

\copyrightnote{The authors. Distributed under a Creative Commons Attribution 4.0 International License}

\begin{center}

{\bf {\large A Special OCNMP Issue in Honour of George W Bluman}}
\end{center}

\smallskip

\smallskip

\Name{On differential equations invariant under a projective transformation group:\\ integrability and reductions}

\Author{Marianna Euler$^{\,1}$, Norbert Euler$^{\,1}$ and Francesco Oliveri$^{\,2}$}

\Address{
$^1$ International Society of Nonlinear Mathematical Physics, 
Auf der Hardt 27,\\
56130 Bad Ems, Germany \&
Centro Internacional de Ciencias, Av. Universidad s/n, Colonia Chamilpa,
 62210 Cuernavaca, Morelos, Mexico\\[0.3cm]
$^2$ Dipartimento di Scienze Matematiche e Informatiche, Scienze Fisiche e Scienze della
Terra, Universit\`a di Messina, Viale F. Stagno d’Alcontres 31, I–98166 Messina, Italy
}

%%%

\Date{Received November 11, 2025; Accepted December 16, 2025}

\setcounter{equation}{0}

\smallskip

\noindent
{\bf Citation format for this Article:}\newline
Marianna Euler, Norbert Euler and Francesco Oliveri,
On differential equations invariant under a projective transformation group: integrability and reductions,
{\it Open Commun. Nonlinear Math. Phys.}, Special Issue:\,Bluman, ocnmp:16901, \pageref{firstpage}--\pageref{lastpage}, 2025.

\strut\hfill

\noindent
{\bf The permanent Digital Object Identifier (DOI) for this Article:}\newline
{\it 10.46298/ocnmp.16901}
\strut\hfill

\begin{abstract}

\noindent 
We consider a projective transformation and establish the invariants for this transformation group up to order seven. We use the obtained invariants to construct a class of nonlinear evolution equations and identify some symmetry-integrable equations in this class. Notably, the only symmetry-integrable evolution equation of order three in this class is a fully-nonlinear equation for which we find the recursion operator and its connection to the Schwarzian KdV. We furthermore establish that higher-order symmetry-integrable equations in this class belong to the hierarchy of the fully-nonlinear 3rd-order equation and prove this for the 5th-order case as well as for the quasilinear 7th-order case. We list all symmetry reductions of this 3rd-order fully-nonlinear symmetry-integrable evolution equation to ordinary differential equations by exploiting the 1-dimensional optimal Lie symmetry subalgebras of the transformation group. We also identify the ordinary differential equations that are invariant under this projective transformation and reduce the order of these equations.

\end{abstract}

\label{firstpage}

%%%% The Article text starts here

\section{Introduction}

In \cite{E-E-76} we reported a class of 3rd-order symmetry-integrable evolution equations of the form
\begin{gather}
u_t=F(u,u_x,u_{xx},u_{3x}),
\end{gather}
which were required to remain invariant under the Möbius (or projective) 
transformation in $u$. In particular we considered the following transformation:
\begin{gather}
\label{Mobius-u}  
{\cal M}: 
\left\{
\ba{l}
\displaystyle{
u({x,t})\mapsto v(\bar{x}, \bar t)= \frac{\alpha_1 u({x,t})+\beta_1}{\alpha_2 u({x,t})+\beta_2}}\\
\\
\displaystyle{{x}\mapsto \bar {x}=x+\epsilon_1  }\\
\\
\displaystyle{{t}\mapsto \bar {t}=t+\epsilon_2}.
\ea
\right.
\end{gather}
Here $\alpha_j,\ \beta_j,\ \gamma_j$, $\delta_j$ and $\epsilon_j$ are real parameters and
\begin{gather*}
%\label{Phi-1}
\Phi=\left(
\ba{cc}
\alpha_1&\beta_1\\
\alpha_2&\beta_2
\ea
\right)
\in SL(2,\mathbb R)\quad \mbox{with $\det \Phi=1$.}
\end{gather*}
Without loss of generality we may set $\beta_2=1$.
The 5-dimensional Lie algebra spanned by the Lie generators 
\begin{gather}
\{Z_1=\pde{\ }{x},\ Z_2=\pde{\ }{u},\ Z_3=u\pde{\ }{u},\ Z_4=u^2\pde{\ }{u},\ Z_5=\pde{\ }{t}\}
\end{gather}
represents this transformation (\ref{Mobius-u}). Using the criteria of symmetry-integrability and invariance under (\ref{Mobius-u}), we obtained the following set of equations \cite{E-E-76}:
 \begin{subequations}
\begin{gather}
\label{SKdV}
u_t=u_xS\ :\quad \mbox{the Schwarzian KdV equation}\\[0.3cm]
\label{full-non-1-intro}
u_t=-2\frac{u_x}{S^{1/2}}\\[0.3cm]
\label{full-non-2-intro}
u_t=\frac{u_x}{(b-S)^2}\\[0.3cm]
%\label{full-non-3-intro}
%u_t=\frac{u_x}{S^2}\\[0.3cm]
\label{full-non-3-intro}
u_t=u_x\left( \frac{a_1-S}{(a_1^2+3a_2)(S^2-2a_1S-3a_2)^{1/2}}\right),
\end{gather}
where $b,\ a_1$ and $a_2$ are arbitrary constants provided that $a_1^2+3a_2\neq 0$. 
Here and throughout this paper we make use of the usual subscript notation to indicate derivatives. In particular, $u_x:=\p u/\p x$, $u_{xx}:=\p^2u/\p x^2$ and $u_{px}:=\p^pu/\p x^p$ for $p\geq 3$.
The symbol $S$ denotes the Schwarzian derivative (see for example \cite{Ovsienko}), which is here defined in terms of $u$ by
\begin{gather}
\label{S-def}
S:=\frac{u_{3x}}{u_x}-\frac{3}{2}\frac{u_{xx}^2}{u_x^2}.
\end{gather}
Note that the -2 in equation (\ref{full-non-1-intro}) is of course not essential, and we could replace it by an arbitrary non-zero constant $\lambda$ or just by 1.
\end{subequations}

Given the fact that
\begin{gather}
\omega_1=\frac{u_t}{u_x}\ \mbox{and}\  \omega_2=S
\end{gather}
are the two fundamental invariants of transformation (\ref{Mobius-u}), it is not surprising that 
equations (\ref{SKdV}) -- (\ref{full-non-3-intro}) are of the general form
\begin{gather}
\label{Psi-Gen}
u_t=u_x\Psi(S),
\end{gather}
where the function $\Psi$ is determined such that (\ref{Psi-Gen}) is symmetry-integrable. This means that 
 (\ref{Psi-Gen}) must admit Lie-Bäcklund symmetries (for details, see for example \cite{Euler-book-2018} and references therein).

It is now natural to investigate the complete Lie point symmetry properties of equations (\ref{SKdV}) -- (\ref{full-non-3-intro}). In doing so, we noticed that equation (\ref{full-non-1-intro}) is the only equation in this class that admits, in addition to the transformation (\ref{Mobius-u}), also a projective transformation in the independent variable $x$. This 
result provided the motivation for our current study.

%\strut\hfill

The paper is organized as follows: In {\bf Section 2} we introduce the projective transformation in the variables $x$ and $u(x,t)$ with a translation in $t$, and calculate all its invariants up to order seven. We then make use of the obtained invariants to construct evolution equations that are invariant under this projective transformation. This leads to four main cases that are distinguished by the order of the invariants used. In {\bf Section 3} we address the problem of identifying symmetry-integrable evolution equations that belong to the mentioned four classes of evolution equations in Section 2. 
We find that there is only one 3rd-order equation in this class that is symmetry-integrable, namely equation 
(\ref{full-non-1-intro}). We derive a recursion operator for this equation and establish the associated hierarchy of symmetry-integrable equations (see Proposition 1 below). Furthermore we prove Proposition 2, which states that 
there exist two fully-nonlinear 5th-order equation that are both invariant under the given projective transformation and symmetry-integrable, as well as one quasilinear equation which belongs to the hierarchy of Proposition 1. 
In Proposition 3 we establish that there exists no symmetry-integrable 6th-order evolution equation in this class, and
in Proposition 4 we state and proof that the 7th-order quasilinear evolution equation that is both invariant under the given transformation and symmetry integrable with a Lie-Bäcklund symmetry of order nine also belongs to the mentioned hierarchy for the 3rd-order equation. In {\bf Section 4} we list all symmetry reductions of the 3rd-order fully-nonlinear symmetry-integrable evolution equation to ordinary differential equations by exploiting the 1-dimensional optimal Lie symmetry subalgebras of the introduced 7 dimensional Lie symmetry algebra.
In {\bf Section 5} we establish the hodograph-type transformation between the 3rd-order symmetry-integrable equation and the Schwarzian Korteweg-de Vries equation, In {\bf Section 6} we restrict the projective transformation to the variables $x$ and $u(x)$ and find its invariants up to order seven. We list all ordinary differential equations (without the fully-nonlinear cases) that are invariant under this projective transformation, solve the 5th-order equation in general and reduce the order of the 6th and 7th-orrder equations. In {\bf Section 7} we make some conclusions.

\section{Invariants and invariant evolution equations}

We consider the following projective transformation in the variables $x$ and $u(x,t)$ with a translation in
 $t$ (see for example \cite{P-E-E-2004}):
\begin{gather}
\label{Mobius-u-x-t}  
{\cal M}_1: 
\left\{
\ba{l}
\displaystyle{
u({x,t})\mapsto v(\bar{x}, \bar t)= \frac{\alpha_1 u({x,t})+\beta_1}{\alpha_2 u({x,t})+\beta_2}}\\
\\
\displaystyle{{x}\mapsto \bar {x}=\frac{\gamma_1 x+\delta_1}{\gamma_2 x+\delta_2}  }\\
\\
\displaystyle{{t}\mapsto \bar {t}=t+\epsilon}.
\ea
\right.
\end{gather}
Here $\alpha_j,\ \beta_j,\ \gamma_j$, $\delta_j$ and $\epsilon$ are real parameters and
\begin{gather*}
%\label{Phi-1}
\Phi_1=\left(
\ba{cc}
\alpha_1&\beta_1\\
\alpha_2&\beta_2
\ea
\right)
\in SL(2,\mathbb R),\quad
%\label{Phi-2}
\Phi_2=\left(
\ba{cc}
\gamma_1&\delta_1\\
\gamma_2&\delta_2
\ea
\right)
\in SL(2,\mathbb R),
\end{gather*}
where $\mbox{det}\, \Phi_1=1$ and  $\mbox{det}\, \Phi_2=1$.
The Lie generators for (\ref{Mobius-u-x-t}) are
\begin{gather}
\label{Lie-Alg-7dim}
\{Z_1=\pde{\ }{x},Z_2=x\pde{\ }{x},Z_3=x^2\pde{\ }{x},
Z_4=\pde{\ }{u},Z_5=u\pde{\ }{u},Z_6=u^2\pde{\ }{u},
%\nn\\[0.3cm]
%\label{Lie-Alg-7dim}
Z_7=\pde{ }{t}\},
\end{gather}
which span the 7-dimensional Lie algebra that
 represents the projective transformation in both $x$ and $u$ with a translation in $t$. 

\strut\hfill

\noindent
{\bf Remark:} Two of the parameters in (\ref{Mobius-u-x-t}) can of course be set to one without loss of generality. For example, we can set $\beta_2=\delta_2=1$.

\strut\hfill

%Indicating the generators in the set (\ref{Lie-Alg-6dim}) by the symbols $Z_j$ and
Applying now the condition
\begin{gather}
\label{cond-Inv-uxt}
Z_j^{(7)} \omega(x,t,u,u_t,u_x,u_{xx},u_{3x},u_{4x},u_{5x},u_{6x},u_{7x})=0, \ j=1,2,\ldots,7,
\end{gather}
we obtain the following four invariants for the seven-dimensional Lie algebra (\ref{Lie-Alg-7dim}) (here $Z_j^{(7)}$ denotes the 7th-order prolongation of the Lie generators $Z_j$ given by (\ref{Lie-Alg-7dim})):
\begin{subequations}
\begin{gather}
\label{invariant-omega-11}
\omega_{1}=\frac{u_t^2}{u_x^2} S\\[0.3cm]
\label{invariant-omega-12}
\omega_{2}=\frac{u_t^4}{u_x^4} \left(
S_{xx}-\frac{5}{4}\frac{S_x^2}{S}+4S^2\right)\\[0.3cm]
\label{invariant-omega-13}
\omega_{3}=\frac{u_t^5}{u_x^5}\left(
S_{3x}-\frac{9}{2}\frac{S_xS_{xx}}{S}+\frac{15}{4}\frac{S_x^3}{S^2}\right)\\[0.3cm]
\label{invariant-omega-14}
\omega_{4}=
\frac{u_t^6}{u_x^6}\left(
S_{4x}-7\frac{S_xS_{3x}}{S}
+\frac{63}{4}\frac{S_x^2S_{xx}}{S^2}
+19SS_{xx}
-\frac{315}{32}\frac{S_x^4}{S^3}
-\frac{95}{4}S_x^2
+34S^3
\right).
\end{gather}
\end{subequations}
We remind that $S$ stands for the Schwarzian derivative (\ref{S-def}) throughout this paper.

\strut\hfill

The most general PDE that is invariant under the transformation (\ref{Mobius-u-x-t}),
albeit with the functional dependence of $\omega$ as indicated in 
(\ref{cond-Inv-uxt}),
is of the form
\begin{gather}
\Psi(\omega_{1},\ \omega_{2}, \ \omega_{3},\ \omega_{4})= 0
\end{gather}
for any given function $\Psi$, where $\omega_{j}$ are the listed invariants. 
In particular, the class of evolution equations that are invariant under (\ref{Mobius-u-x-t}) can be presented in four main cases according to their orders, where we refer to the invariants  (\ref{invariant-omega-11}) -- (\ref{invariant-omega-14}) listed above. For convenience we define the following three variables in terms of $S$ and its $x$-derivatives:
\begin{subequations}
\begin{gather}
\label{Omega-1}
\Omega_1:=\frac{\omega_2}{\omega_1^2}
=\frac{S_{xx}}{S^2}-\frac{5}{4}\frac{S_x^2}{S^3}+4\\[0.3cm]
\label{Omega-2}
\Omega_2:=\frac{\omega_3}{\omega_1^{5/2}}
=\frac{S_{3x}}{S^{5/2}}
-\frac{9}{2}\frac{S_xS_{xx}}{S^{7/2}}+\frac{15}{4}\frac{S_x^3}{S^{9/2}}\\[0.3cm]
\label{Omega-3}
\Omega_3:=\frac{\omega_4}{\omega_1^3}
=\frac{S_{4x}}{S^3}
-7\frac{S_xS_{3x}}{S^4}
+\frac{63}{4}\frac{S_x^2S_{xx}}{S^5}
+19\frac{S_{xx}}{S^2}
-\frac{315}{32}\frac{S_x^4}{S^6}
-\frac{95}{4}\frac{S_x^2}{S^3}
+34.
\end{gather}
\end{subequations}
We note that the relations between $\Omega_1$, $\Omega_1$ and $\Omega_3$ are as follows:
\begin{subequations}
\begin{gather}
\Omega_2=\frac{\Omega_{1x}}{\sqrt{S}}\\[0.3cm]
\Omega_3=\frac{\Omega_2\Omega_{2x}}{\Omega_{1x}}
+\frac{9}{2}\Omega_1^2-17\Omega_1+30.
\end{gather}
\end{subequations}

\strut\hfill

\noindent
{\bf Case 1.1:} {\it The third-order case.} There is only one invariant available here, $\omega_1$, so we consider
\begin{gather}
\omega_1=\lambda^2
\end{gather}
which gives the fully-nonlinear 3rd-order evolution equation 
\begin{gather}
\label{Eq-Case1.1}
u_t=\lambda \frac{u_x}{S^{1/2}}.
\end{gather} 
Here $\lambda$ is an arbitrary non-zero constant.

\strut\hfill

\noindent
{\bf Case 1.2:} {\it The fifth-order case.} We have two invariants available, namely $\omega_1$ and $\omega_2$, which provides the general case
\begin{gather}
\omega_1^{1/2}=F(\Omega_1)
\end{gather}
with $F$ any given function for which
\begin{gather*}
\frac{dF}{d\Omega_1}\neq 0.
\end{gather*}
This leads to the following 5th-order evolution equation:
\begin{gather}
\label{Eq-1.2a}
u_t=\frac{u_x}{S^{1/2}}\,F(\Omega_1).
\end{gather}
Note that the relation
\begin{gather*}
\omega_2^{1/4}=\tilde F(\Omega_1)
\end{gather*}
for some arbitrary function $\tilde F$ leads to the same equation (\ref{Eq-1.2a}).

\strut\hfill

\noindent
{\bf Case 1.3:} {\it The sixth-order case.} We have three invariants available, namely $\omega_1$, $\omega_2$
and $\omega_3$, which provides the general case
\begin{gather}
\omega_1^{1/2}=F(\Omega_1,\Omega_2)
\end{gather}
with $F$ any given function for which
\begin{gather*}
\pde{F}{\Omega_2}\neq 0.
\end{gather*}
This leads to the following 6th-order evolution equation:
\begin{gather}
\label{Eq-Case1.3a}
u_t=\frac{u_x}{S^{1/2}}\,F(\Omega_1,\Omega_2).
\end{gather}
Note that the relations
\begin{gather*}
\omega_2^{1/4}=\tilde F_1(\Omega_1,\Omega_2)
\end{gather*}
and 
\begin{gather*}
\omega_3^{1/5}=\tilde F_2(\Omega_1,\Omega_2)
\end{gather*}
lead, for arbitrary functions $\tilde F_1$ and $\tilde F_2$, to the same equation (\ref{Eq-Case1.3a}).

\strut\hfill

\noindent
{\bf Case 1.4:} {\it The seventh-order case.} We have four invariants available, namely $\omega_1$, $\omega_2$,
$\omega_3$ and $\omega_4$,  which provides the general case
\begin{gather}
\omega_1^{1/2}=F(\Omega_1,\Omega_2, \Omega_3)
\end{gather}
with $F$ any given function for which
\begin{gather*}
\pde{F}{\Omega_3}\neq 0.
\end{gather*}
This leads to the following 7th-order evolution equation:
\begin{gather}
\label{Eq-Case1.4a}
u_t=\frac{u_x}{S^{1/2}}\,F(\Omega_1,\Omega_2, \Omega_3).
\end{gather}
Note that the relations
\begin{gather*}
\omega_2^{1/4}=\tilde F_1(\Omega_1,\Omega_2, \Omega_3),\quad
\omega_3^{1/5}=\tilde F_2(\Omega_1,\Omega_2, \Omega_3)
\end{gather*}
and 
\begin{gather*}
\omega_4^{1/6}=\tilde F_3(\Omega_1,\Omega_2, \Omega_3)
\end{gather*}
lead, for arbitrary functions $\tilde F_1$, $\tilde F_2$ and $\tilde F_3$, to the same equation (\ref{Eq-Case1.4a}).

\section{Symmetry-integrable evolution equations}

The aim here is to find evolution equation that are both invariant under the transformation (\ref{Mobius-u-x-t})
and {\it symmetry-integrable}. By symmetry-integrable equations we mean equations that admit higher-order generalised symmetries, the so-called Lie-Bäcklund symmetries, and a recursion operator that generate these symmetries. This then typically leads to a hierarchy of higher-order symmetry-integrable equations (see for example \cite{Euler-book-2018} and the references therein). 

\subsection{The 3rd-order evolution equation (\ref{Eq-Case1.1})}
The general 3rd-order evolution equation invariant under (\ref{Mobius-u-x-t}) is given by equation 
(\ref{Eq-Case1.1}) in Case 1.1 {\it viz.}
\begin{gather*}
%\label{FN-3rd-order-Eq-S-1}
u_t=\lambda \frac{u_x}{\sqrt{S}}.
\end{gather*}
We find that (\ref{Eq-Case1.1}) admits a 5th-order Lie-Bäcklund symmetry and the following 
2nd-order recursion operator 
\begin{gather}
\label{Recursion-op-1}
R[u]=G_2D_x^2+G_1D_x+G_0+\lambda \frac{u_x}{\sqrt{S}}\,D_x^{-1}
\circ \Lambda,
\end{gather}
where
\begin{subequations}
\begin{gather}
\label{G0}
G_0=\frac{k_2}{4}\left(\frac{S_{xx}}{S^2}
+\frac{u_{xx}}{u_x}\frac{S_x}{S^2}
-\frac{3}{2}\frac{S_x^2}{S^3}
-\frac{u_{xx}^2}{u_x^2}\frac{1}{S}\right)
+k_1+k_2
\\[0.3cm]
G_1=k_2\left(\frac{u_{xx}}{u_x}\frac{1}{S}-\frac{1}{4}\frac{S_x}{S^2}\right)\\[0.3cm]
G_2=-\frac{k_2}{2S}\\[0.3cm]
\label{lam}
\Lambda=-\frac{k_2}{4\lambda u_x}\left(\frac{S_{3x}}{S^{3/2}}-\frac{9}{2}\frac{S_xS_{xx}}{S^{5/2}}
+\frac{15}{4}\frac{S_x^3}{S^{7/2}}\right).
\end{gather}
\end{subequations}
Here $k_1$ and $k_2$ are arbitrary constants and $k_2\neq 0$.  
This establishes that (\ref{Eq-Case1.1}) is symmetry-integrable, which is in agreement with the result 
reported in \cite{E-E-76}. This leads to

%%&&&&&&&&&&&&&&&&&&&&&&&&&&&&&&&&&&&&&&&&&&&

\strut\hfill

\begin{proposition}
\label{Prop-R-for-3rd-order}
The fully-nonlinear equation  (\ref{Eq-Case1.1}), viz.
\begin{gather*}
u_t=\lambda\frac{u_x}{\sqrt{S}}
\end{gather*}
provides the following symmetry-integrable hierarchy:
\begin{gather}
\label{hierarchy-1}
u_{t_j}=R^j[u] \,u_t,\qquad j=1,2,\ldots,
\end{gather}
whereby every member of the hierarchy (\ref{hierarchy-1}) is of the form 
\begin{gather}
u_{t_j}=u_x\Psi_j(S,S_x,S_{xx},\ldots,S_{nx})
\end{gather}
and
$R[u]$ is the recursion operator (\ref{Recursion-op-1}). 

\end{proposition}

\strut\hfill

\noindent
The second member of the hierarchy (\ref{hierarchy-1}) leads the 5th-order symmetry-integrable equation
\begin{gather}
\label{Compare}
u_{t_1}=R[u]\, u_{t}=\lambda u_x\left[ \frac{k_2}{4}\frac{S_{xx}}{S^{5/2}}
-\frac{5k_2}{16}\frac{S_x^2}{S^{7/2}}
+\left(\frac{k_2}{2}+k_1\right)\frac{1}{S^{1/2}}    \right]
\end{gather}
or, without loss of generality we let $k_2=1$, we have
\begin{gather}
\label{Compare-No-k2}
u_{t_1}=R[u]\, u_{t}=\lambda u_x\left[ \frac{1}{4}\frac{S_{xx}}{S^{5/2}}
-\frac{5}{16}\frac{S_x^2}{S^{7/2}}
+\left(\frac{1}{2}+k_1\right)\frac{1}{S^{1/2}}    \right].
\end{gather}
The third member of the hierarchy (\ref{hierarchy-1}) gives the 7th-order symmetry-integrable equation (with $k_2=1$)
\begin{gather}
%\label{Compare-2}
u_{t_2}=
%R[u]^2\, u_{t}\nn\\[0.3cm]
\lambda u_x\left[
-\frac{1}{8}\frac{S_{4x}}{S^{7/2}}
+\frac{7}{8}\frac{S_xS_{3x}}{S^{9/2}}
+\frac{21}{32}\frac{S_{xx}^2}{S^{9/2}}
-\frac{231}{64}\frac{S_x^2S_{xx}}{S^{11/2}}
+\frac{1}{4}\left(2k_1+1\right)\frac{S_{xx}}{S^{5/2}}
\right.\nn\\[0.3cm]
\label{7th-order-EE}
\quad
%+\left(\frac{k_2^2}{4}+\frac{k_1k_2}{2}\right)\frac{S_{xx}}{S^{5/2}}
+\frac{1155}{64}\frac{S_x^4}{S^{13/2}}
-\frac{5}{16}\left(2k_1+1\right)\frac{S_x^2}{S^{7/2}}
%\nn\\[0.3cm]
%\label{7th-order-EE}
\left.
+\frac{1}{4}\left(2k_1+1\right)^2\frac{1}{S^{1/2}}\right].
\end{gather}

\smallskip

\noindent
{\bf Note:} {\it For $k_1=-1$, $k_2=-2$ and $\lambda=-2$, equation (\ref{Compare}) is identical to equation (2.10a) in  
 \cite{E-E-78}, 
so this result coincides with the hierarchy that was generated from the equations in terms of the Schwarzian derivative $S$ reported in \cite{E-E-78}.
}

\subsection{The 5th-order evolution equations (\ref{Eq-1.2a}) }

For symmetry-integrable 5th-order evolution equations
%to admit higher-order Lie-Bäcklund symmetries, hence for symmetry-integrability, 
the following statement is useful:

\strut\hfill

\noindent
{\bf Lemma 1:} 
{\it
If a 5th-order evolution equation of the form
\begin{gather}
\label{Prop-NC-Symm-5th-EQ}
u_t=u_x\Psi(x,t,S,S_x,S_{xx})
\end{gather}
is symmetry-integrable for a given function $\Psi$ then this function must satisfy the following condition:
\begin{gather}
\label{Prop-NC-Symm-5th}
5\pde{\Psi}{S_{xx}}\,\frac{\p^3 \Psi}{\p S_{xx}^3}-8\left(\frac{\p^2 \Psi}{\p S_{xx}^2}\right)^2=0.
\end{gather}
}

\strut\hfill

\noindent
{\bf Proof:}
We assume that (\ref{Prop-NC-Symm-5th-EQ}) is symmetry-integrable. 
By definition the equation must therefore admit Lie-Bäcklund symmeties of order $n$, where $n$ is an integer larger than five. Conditions (\ref{Prop-NC-Symm-5th})  then
follows directly from the Lie-Bäcklund symmetry invariance condition 
\begin{gather}
\label{IC}
\left.
\vphantom{\frac{DA}{DB}}
L_E[u]Q\right|_{E=0}=0,
\end{gather}
where $E$ denotes the equation, i.e. $E:=u_t-u_x\Psi $, and $Q=Q(x,t,u_x,u_{xx},\ldots ,u_{nx})$ is the characteristic of the Lie-Bäcklund symmetry generator 
\begin{gather}
\label{IC-LB}
Z_{LB}=Q(x,t,u_x,u_{xx},\ldots ,u_{nx})\pde{\ }{u}.
\end{gather}
Here $L_E[u]$ denotes the linear operator
\begin{gather}
L_E[u]:=\pde{E}{u}+\pde{E}{u_t}D_t+\pde{E}{u_x}D_x+\pde{E}{u_{xx}}D_x^2+\cdots
+\pde{E}{u_{qx}}D_x^q
\end{gather}
(in this case $q=5$) and $S$ is the Schwarzian derivative (\ref{S-def}).
\strut\hfill $\Box$

\strut\hfill

\noindent
By studying the Lie-Bäcklund symmetry properties of the two 5th-order equations (\ref{Eq-1.2a}) given in Case 1.2, we are led to the following
%
%We need to study the two equations given in Case 1.2 to establish for which functions $F(\Omega_1)$
%the equations admit Lie-Bäcklund symmetries of order 7 and order 11. This leads to

\strut\hfill

\begin{proposition}
The only fully-nonlinear 5th-order evolution equations that are both invariant under the projective transformation (\ref{Mobius-u-x-t}) and symmetry-integrable are the following two equations:
\begin{subequations}
\begin{gather}
\label{5th-FNE-1}
u_t=u_xS^{5/6}\left(
S_{xx}-\frac{5}{4}\frac{S_x^2}{S}+\frac{2}{3}S^2\right)^{-2/3}\\[0.3cm]
\label{5th-FNE-2}
u_t=u_xS^{5/6}\left(
S_{xx}-\frac{5}{4}\frac{S_x^2}{S}-\frac{8}{3}S^2\right)^{-2/3}.
\end{gather}
\end{subequations}
Furthermore, the only quasilinear 5th-order evolution equation that is both invariant under the projective transformation (\ref{Mobius-u-x-t}) and symmetry-integrable is
equation (\ref{Compare-No-k2}), namely the second member of the hierarchy 
(\ref{hierarchy-1}) of Proposition 1
\begin{gather*}
%\label{Compare-No-k2}
u_{t}=\lambda u_x\left[ \frac{1}{4}\frac{S_{xx}}{S^{5/2}}
-\frac{5}{16}\frac{S_x^2}{S^{7/2}}
+\left(\frac{1}{2}+k_1\right)\frac{1}{S^{1/2}}    \right].
\end{gather*}

\end{proposition}

\smallskip

\noindent
{\bf Proof:} 
By {\bf Case 1.2} we need to consider 
equation (\ref{Eq-1.2a}), {\it viz.}
\begin{gather*}
%\label{Eq-1.2a}
u_t=\frac{u_x}{S^{1/2}}\,F_1(\Omega_1),
\end{gather*}
where $\Omega_1$ is given by (\ref{Omega-1}). Condition (\ref{Prop-NC-Symm-5th}) with
\begin{gather}
\Psi=\frac{1}{S^{1/2}}\,F_1(\Omega_1)
\end{gather}
leads to the following condition on $F_1$:
\begin{gather}
\label{Prop-NC-Symm-F1}
5\frac{dF_1}{d\Omega_1}\,\frac{d^3 F_1}{d \Omega_1^3}-8
\left(\frac{d^2 F_1}{d \Omega_1^2}\right)^2=0.
\end{gather}
Clearly $\displaystyle{\frac{d^2 F_1}{d \Omega_1^2}=0}$ satisfies (\ref{Prop-NC-Symm-F1}), so that (\ref{Eq-1.2a}) takes the form
\begin{gather}
\label{Case 1.2a-conclude}
u_t=\frac{u_x}{S^{1/2}}\left[c_1\left( \frac{S_{xx}}{S^2}-\frac{5}{4}\frac{S_x^2}{S^3}+4\right) +c_2\right],
\end{gather}
where $c_1$ and $c_2$ are arbitrary constants. Equation (\ref{Case 1.2a-conclude}) is the linear combination of the first two equations in the hierarchy (\ref{hierarchy-1}) of Proposition 1, and therefore symmetry-integrable. Furthermore, the general solution of (\ref{Prop-NC-Symm-F1}) is
\begin{gather}
F_1(\Omega_1)=c_1\left(\Omega_1+c_2\right)^{-2/3}+c_3,
\end{gather}
where $c_1,\ c_2$ and $c_3$ are arbitrary constants. This leads to the following fully-nonlinear evolution equation
\begin{gather}
\label{Case-1.2a-NI}
u_t=\frac{u_x}{S^{1/2}}\left[c_1\left(\frac{S_{xx}}{S^2}-\frac{5}{4}\frac{S_x^2}{S^3}+4
+c_2\right)^{-2/3}+c_3  \right].
\end{gather}
Applying the invariance condition 
(\ref{IC})  we find that Lie-Bäcklund symmetries up to order 11 exist for equation 
(\ref{Case-1.2a-NI}) for the case $c_1=1$, $c_2=-10/3$ or $c_2=-20/3$, and $c_3=0$, which corresponds to equation (\ref{5th-FNE-1}) and
equation (\ref{5th-FNE-2}), respectively.
\strut\hfill $\Box$

\subsection{The 6th-order evolution equation (\ref{Eq-Case1.3a}) } 

For symmetry-integrable 6th-order evolution equations
%to admit higher-order Lie-Bäcklund symmetries, hence for symmetry-integrability, 
the following statement was obtained by applying the Lie-Bäcklund symmetry invariance condition
(\ref{IC}) for symmetries of order $n\geq 7$:

\strut\hfill

\noindent
{\bf Lemma 2:} 
{\it
Consider the general 6th-order evolution equation
\begin{gather}
\label{Gen-6th-order}
u_t=\Psi(x,t,u,u_x,u_{xx},\ldots,u_{6x}).
\end{gather}
Then the following necessary conditions for symmetry-integrability of (\ref{Gen-6th-order}) apply:
\begin{enumerate}
\item
The condition
\begin{gather}
\label{Prop-NC-Symm-6th}
\frac{\p^2 \Psi\ }{\p u_{6x}^2}=0
\end{gather}
must hold. That is (\ref{Gen-6th-order}) must be of the form
\begin{gather}
\label{Gen-6th-lin}
u_t=\Psi_1(x,t,u,u_x,\ldots, u_{5x})u_{6x}+\Psi_2(x,t,u,u_x,\ldots, u_{5x})
\end{gather}
for some functions $\Psi_1$ and $\Psi_2$.
\item
For equation (\ref{Gen-6th-lin}), the following condition on the function $\Psi_1$ must hold:
\begin{gather}
\label{cond-Psi_1}
\pde{\Psi_1}{u_{kx}}=0,\ \mbox{for}\  k\in\{3,4,5\}.
\end{gather}

\end{enumerate}
That is, any 6th-order symmetry-integrable evolution equation must be linear in $u_{6x}$, and the coefficient of 
$u_{6x}$ can not depend on $u_{3x},\ u_{4x}$ or $u_{5x}$.

}

\strut\hfill

\noindent

\noindent
By Case 1.3 and Lemma 2 we obtain

\strut\hfill

\noindent
\begin{proposition}
There exists no 6th-order evolution equation that is both invariant under the transformation (\ref{Mobius-u-x-t}) and
symmetry-integrable.

\end{proposition}

\strut\hfill

\noindent
{\bf Proof:}
We know from Case 1.3 that any 6th-order evolution equation that is invariant under (\ref{Mobius-u-x-t}) is of the form (\ref{Eq-Case1.3a}), viz.
\begin{gather*}
u_t=\frac{u_x}{S^{1/2}}F(\Omega_1,\Omega_2),
\end{gather*}
where $\Omega_1$ and $\Omega_2$ are given by (\ref{Omega-1}) and (\ref{Omega-2}), respectively. By Lemma 2 it follows that (\ref{Eq-Case1.3a}) must be of the form
\begin{gather}
\label{6th-linear}
u_t=\frac{u_x}{S^{1/2}}\left[
\vphantom{\frac{DA}{DB}}
f_1(\Omega_1)\Omega_2+f_2(\Omega_1)\right]
\end{gather}
for some functions $f_1$ and $f_2$. In particular, equation (\ref{6th-linear}) is of the form
\begin{gather}
u_t=\frac{f_1(\Omega_1)}{S^3}\,u_{6x}+\phi(u,u_x,u_{xx},\ldots,u_{5x})
\end{gather}
and by condition (\ref{cond-Psi_1}) it follows that $f_1=0$.\strut\hfill $\Box$

\subsection{The 7th-order evolution equation (\ref{Eq-Case1.4a})}

Applying Conjecture 1 as stated in \cite{E-E-2025-v5}, we can state the following for the 7th-order equation 
(\ref{Eq-Case1.4a}), viz.
\begin{gather*}
%\label{Eq-Case1.4a}
u_t=\frac{u_x}{S^{1/2}}\,F(\Omega_1,\Omega_2, \Omega_3):
\end{gather*}

\strut\hfill

\noindent
{\bf Lemma 3:} {\it Consider the general 7th-order evolution equation
\begin{gather}
\label{Gen-7th-order}
u_t=\Phi(x,t,u,u_x,u_{xx},\ldots,u_{7x}).
\end{gather}
\begin{enumerate}

\item
A necessary condition for symmetry-integrability of (\ref{Gen-7th-order}) is given by the following condition:
\begin{gather}
\label{7th-order-necess-cond}
7\pde{\Phi}{u_{7x}}\frac{\p^3 \Phi}{\p u_{7x}^3}-11\left(
\pdd{\Phi}{u_{7x}}\right)^2=0.
\end{gather}

\item
The necessary condition (\ref{7th-order-necess-cond}) applied to equation (\ref{Eq-Case1.4a}), viz.
\begin{gather*}
%\label{Eq-Case1.4a}
u_t=\frac{u_x}{S^{1/2}}\,F(\Omega_1,\Omega_2, \Omega_3),
\end{gather*}
provides two subcases for $F$ for the symmetry-integrability of equation (\ref{Eq-Case1.4a}), namely 
\begin{subequations}
\begin{gather}
\label{Quasi-lin-7th-order-F}
F(\Omega_1,\Omega_2,\Omega_3)=\Psi_{11}(\Omega_1,\Omega_2)\,\Omega_3+\Psi_{12}(\Omega_1,\Omega_2)
\end{gather}
for some functions $\Psi_{11}$ and $\Psi_{12}$, or
\begin{gather}
\label{Full-NL-7th-order-F}
F(\Omega_1,\Omega_2,\Omega_3)=
\frac{\Psi_{21}(\Omega_1,\Omega_2)}{\left[
\vphantom{\frac{da}{db}}
\Omega_3+\Psi_{22}(\Omega_1,\Omega_2)\right]^{3/4}}
+\Psi_{23}(\Omega_1,\Omega_2)
\end{gather}
for some functions $\Psi_{21},\ \Psi_{22}$ and $\Psi_{23}$. Here $\Omega_1,\ \Omega_2$ and $\Omega_3$ are given by
(\ref{Omega-1}), (\ref{Omega-2}) and (\ref{Omega-3}), respectively.

\end{subequations}

\end{enumerate}
}

\strut\hfill

\noindent
For 7th-order quasilinear evolutions we have the following 

\strut\hfill

\noindent
{\bf Lemma 4:} {\it Consider the general 7th-order quasilinear evolution equation
\begin{gather}
\label{7th-Quasi-L-Gen}
u_t=\Phi_1(x,t,u,u_{x},u_{xx},\ldots,u_{6x}) u_{7x}+\Phi_2(x,t,u,u_{x},u_{xx},\ldots,u_{6x}).
\end{gather}

\begin{enumerate}

\item
If (\ref{7th-Quasi-L-Gen}) is symmetry-integrable with a lowest-order Lie-Bäcklund symmetry of order eight, 
then $\Phi_1$ is constrained by the following conditions:

\begin{gather}
\label{lemma4-cond-1}
\pde{\Phi_1}{u_{kx}}=0\ \mbox{for}\ k=3,4,5,6.
\end{gather}

\item
If (\ref{7th-Quasi-L-Gen}) is symmetry-integrable with a lowest-order Lie-Bäcklund symmetry of order nine, then $\Phi_1$ is constrained by the following conditions:
\begin{subequations}
\begin{gather}
\label{Lemma4-cond-2}
\pde{\Phi_1}{u_{kx}}=0\ \mbox{for}\ k=4,5,6
\end{gather}
and
\begin{gather}
\label{7th-order-QL-Cond}
\frac{7}{9}\Phi_1^2\,\frac{\p^3 \Phi_1}{\p u_{3x}^3}
-3\Phi_1\,\pde{\Phi_1}{u_{3x}}\, \pdd{\Phi_1}{u_{3x}}
+\frac{16}{7}\left(
\pde{\Phi_1}{u_{3x}}\right)^3=0.
\end{gather}
\end{subequations}

The general solution of (\ref{7th-order-QL-Cond}) is
\begin{gather}
\Phi_1(x,t,u,u_x,u_{xx},u_{3x})=
\frac{\phi_1}{\left[(u_{3x}+\phi_2)^2+\phi_3\right]^{7/2}},
\end{gather}
where $\phi_j=\phi_j(x,t,u,u_x,u_{xx})$, $j=1,2,3$, are arbitrary functions.

\end{enumerate}

}

\strut\hfill

\noindent
Both Lemma 3 and Lemma 4 can be verified by applying the Lie-Bäcklund symmetry invariance condition (\ref{IC}).

\strut\hfill

\noindent
Using Lemma 3 and Lemma 4 we can state the following for the quasilinear 7th-order case:

\strut\hfill

\begin{proposition}
The only 7th-order quasilinear evolution equation that is both invariant under the transformation (\ref{Mobius-u-x-t}) and symmetry-integrable with a Lie-Bäcklund symmetry of order nine,
is the 7th-order equation that belongs to the hierarchy (\ref{hierarchy-1}) given in Proposition 1, namely the equation 
(\ref{7th-order-EE}), viz
\begin{gather*}
%\label{Compare-2}
u_{t}=
%R[u]^2\, u_{t}\nn\\[0.3cm]
\lambda u_x\left[
-\frac{1}{8}\frac{S_{4x}}{S^{7/2}}
+\frac{7}{8}\frac{S_xS_{3x}}{S^{9/2}}
+\frac{21}{32}\frac{S_{xx}^2}{S^{9/2}}
-\frac{231}{64}\frac{S_x^2S_{xx}}{S^{11/2}}
+\frac{1}{4}\left(2k_1+1\right)\frac{S_{xx}}{S^{5/2}}
\right.\nn\\[0.3cm]
%\label{7th-order-EE}
%
\quad
%+\left(\frac{k_2^2}{4}+\frac{k_1k_2}{2}\right)\frac{S_{xx}}{S^{5/2}}
+\frac{1155}{64}\frac{S_x^4}{S^{13/2}}
-\frac{5}{16}\left(2k_1+1\right)\frac{S_x^2}{S^{7/2}}
%\nn\\[0.3cm]
%\label{7th-order-EE}
\left.
+\frac{1}{4}\left(2k_1+1\right)^2\frac{1}{S^{1/2}}\right].
\end{gather*}

\end{proposition}

\strut\hfill

\noindent
{\bf Proof:} By Lemma 3 we need to consider the 7th-order quasilinear equation 
\begin{gather}
\label{7th-order-SIE}
u_t=\frac{u_x}{S^{1/2}}\left[
\vphantom{\frac{DA}{DB}}
\Psi_{11}(\Omega_1,\Omega_2)\Omega_3+\Psi_{12}(\Omega_1,\Omega_2)
\right].
\end{gather}
In particular, we have
\begin{gather}
u_t=\Psi_{11}(\Omega_1,\Omega_2)\,\frac{u_{7x}}{S^{7/2}}+\cdots.
\end{gather}
We recall that $\Omega_1$ contains the fifth derivative $u_{5x}$ and $\Omega_2$ the sixth derivative $u_{6x}$. By Lemma 4, it follows that equation (\ref{7th-order-SIE}) can not admit an 8th-order Lie-Bäcklund symmetry since
\begin{gather}
\label{Phi_1-proof-Prop4}
\Phi_1=\frac{\Psi_{11}}{S^{7/2}}
\end{gather}
leads to $\Psi_{11}=0$ by condition (\ref{lemma4-cond-1}). However, both condition (\ref{Lemma4-cond-2}) and condition 
(\ref{7th-order-QL-Cond}) are satisfied for $\Phi_1$ given by (\ref{Phi_1-proof-Prop4}) for a 9th-order Lie-Bäcklund symmetry, whereby $\Psi_{11}$ is any non-zero constant. Appling the Lie-Bäcklund symmetry invariance condition (\ref{IC}) it follows that (\ref{7th-order-EE}) is the only equation in this class that admits a 9th-order Lie-Bäcklund symmetry. \strut\hfill$\Box$

\strut\hfill

\noindent
Regarding the fully-nonlinear 7th-order case (\ref{Eq-Case1.4a}):
We were not able to establish any symmetry-integrable case for the equations (\ref{Eq-Case1.4a}) with $F$ given by 
(\ref{Full-NL-7th-order-F}), i.e. the fully-nonlinear 7th-order equation
\begin{gather}
\label{7th-order-FN-Q}
u_t=\frac{u_x}{S^{1/2}}\frac{\Psi_{21}(\Omega_1,\Omega_2)}{\left[
\vphantom{\frac{da}{db}}
\Omega_3+\Psi_{22}(\Omega_1,\Omega_2)\right]^{3/4}}
+\Psi_{23}(\Omega_1,\Omega_2).
\end{gather}
The symmetry-integrability of (\ref{7th-order-FN-Q}), or its symmetry-nonintegrability, is therefore, at this point in time, an open problem. 
We should mention that in a recent study \cite{E-E-2025-v5} we established that the fully-nonlinear equation
\begin{gather}
u_t=u_{7x}^{-3/4}
\end{gather}
is not symmetry-integrable. We therefore do not expect the 7th-order equation (\ref{7th-order-FN-Q}) to be symmetry-integrable for any choice of the functions $\Psi_{21}$, $\Psi_{22}$ or $\Psi_{23}$, but at this point we are not able to prove this.

%%%%%%%%%%%%%%%%%%%%%%%%%%%%%%
%%%%%%%%%%%%%%%%%%%%%%%%%%

\section{Symmetry reductions}
%In this section we systematically reduce the 3rd-order symmetry-integrable evolution equation (\ref{Eq-Case1.1}), viz.
%\begin{gather*}
%%\label{Eq-Case1.1}
%u_t=\lambda \frac{u_x}{S^{1/2}}
%\end{gather*} 
%to ordinary differential equations using the one dimensional optimal subalgebras of the seven dimensional Lie algebra spanned by the generators (\ref{Lie-Alg-7dim}). 
%
%Using the program {\it SymboLie} \cite{Amata-Oliveri-Sgroi-JGP-2024} (see also \cite{Amata-Oliveri-Sgroi-2024}) we obtain the following 27 one dimensional optimal subalgebras for the 7-dimensional Lie symmetry algebra spanned by (\ref{Lie-Alg-7dim}):

In this section, looking for group invariant solutions, we systematically reduce the 3rd-order symmetry-integrable evolution equation (\ref{Eq-Case1.1}), {\it viz.}
\begin{gather*}
%\label{Eq-Case1.1}
u_t=\lambda \frac{u_x}{S^{1/2}}
\end{gather*} 
to ordinary differential equations. This is accomplished by analyzing the different Lie subgroups of the admitted symmetries. Different Lie subgroups of symmetries in principle lead to different invariant solutions; nevertheless, subgroups can be partitioned into conjugacy classes, so that equivalent subgroups determine equivalent invariant solutions \cite{Ovsiannikov,Olver} linked to each other by the action of some subgroup. The intimate connection between Lie groups and Lie algebras allows us to identify inequivalent subgroups of a Lie group by classifying inequivalent Lie subalgebras; the latter problem can be addressed using the inner automorphisms of the Lie algebra \cite{Ovsiannikov,Olver}. A set of representatives of the inequivalent classes of Lie subalgebras is called an optimal system. The construction of an optimal system of Lie subalgebras can be done using a computer algebra program like {\it SymboLie} \cite{Amata-Oliveri-Sgroi-JGP-2024} (see also \cite{Amata-Oliveri-Sgroi-2024} and 
\cite{Amata-Oliveri-Sgroi-OCNMP-2025}). Therefore,  
using {\it SymboLie}, the one dimensional optimal subalgebras of the seven dimensional Lie algebra spanned by the generators (\ref{Lie-Alg-7dim}) can be determined. 
As a result, we obtain the following 27 one-dimensional optimal subalgebras for the 7-dimensional Lie symmetry algebra spanned by (\ref{Lie-Alg-7dim}):
\begin{gather}
\{Z_1\}, \{Z_2\}, \{Z_4\}, \{Z_5\},\{Z_7\},\nn\\[0.2cm]
\{Z_1+\alpha_1 Z_3\}, \{Z_1+\alpha_1 Z_4\}, \{Z_2+\alpha_1 Z_4\}, \{Z_1+\alpha_1 Z_5\}, 
\{Z_2+a_1 Z_5\}, \{Z_4+\alpha_1 Z_6\},\nn\\
 \{Z_1+\alpha_1 Z_7\}, 
 \{Z_2+a_1 Z_7\}, 
 \{Z_4+\alpha_1 Z_7\}, 
 \{Z_5+a_1 Z_7\}, \nn\\[0.2cm]
 \{Z_1+\alpha_1Z_3+\alpha_2 Z_4\},
 \{Z_1+\alpha_1Z_4+\alpha_2 Z_6\},
 \{Z_1+\alpha_1Z_3+a_1 Z_7\},
 \{Z_1+\alpha_1Z_4+a_1 Z_7\},\nn\\
 \{Z_2+\alpha_1Z_4+a_1 Z_7\},
 \{Z_1+\alpha_1Z_5+a_1 Z_7\},
 \{Z_2+a_1Z_5+a_2 Z_7\},
 \{Z_4+\alpha_1Z_6+a_1 Z_7\},\nn\\[0.2cm]
 \{Z_1+\alpha_1 Z_3+\alpha_2 Z_4+a_1Z_6\},
 \{Z_1+\alpha_1 Z_3+\alpha_2 Z_4+a_1Z_7\},
 \{Z_1+\alpha_1 Z_4+\alpha_2 Z_6+a_1 Z_7\}\nn\\[0.2cm]
 \label{1D-SubAlg}
 \{Z_1+\alpha_1 Z_3+\alpha_2 Z_4+a_1 Z_6+a_2 Z_7\},
\end{gather}
where $a_k\in \mathbb R\backslash\{0\}$ and $\alpha_k=\pm 1$ for $k=1,2$. We apply the given subalgebras and their corresponding symmetry Ansatz that follow from the general solutions of the invariant surfaces condition 
$Q=0$, where
\begin{gather}
Q=\xi_1(x,t,u)u_x+\xi_2(x,t,u)u_t-\eta(x,t,u)
\end{gather}
for the symmetry generator
\begin{gather}
Z=\xi_1(x,t,u)\pde{\ }{x}+\xi_2(x,t,u)\pde{\ }{t}+\eta(x,t,u)\pde{\ }{u}.
\end{gather}
This procedure is well know (see for example \cite{Bluman-Kumei} or \cite{Olver}).
Below we list all sensible and nontrivial reductions of the 3rd-order evolution equaton (\ref{Eq-Case1.1}) to corresponding ordinary differential equations (ODEs) using the given subalgebras (\ref{1D-SubAlg}):

\strut\hfill

\noindent
{\bf 4.1 The subalgebra spanned by $\{Z_1+\alpha_1 Z_4\}$} leads to the symmetry Ansatz
\begin{subequations}
\begin{gather}
u(x,t)=V(\omega)+\alpha_1 \ln(x),\quad x>0\\[0.3cm]
\omega(x,t)=t.
\end{gather}
\end{subequations}
The reduced ODE for equation (\ref{Eq-Case1.1})
then takes the form
\begin{gather}
V_\omega=\sqrt{2}\lambda\alpha_1
\end{gather}
with the general solution
\begin{gather}
V(\omega)=\sqrt{2}\lambda\alpha_1 \omega +C_1,
\end{gather}
where $C_1$ is a constant of integration. A special solution of (\ref{Eq-Case1.1}) is then
\begin{gather}
u(x,t)=\alpha_1\ln(x)+\sqrt{2}\lambda \alpha_1 t+C_1.
\end{gather}

\strut\hfill

\noindent
{\bf 4.2 The subalgebra spanned by $\{Z_1+\alpha_1 Z_5\}$} leads to the symmetry Ansatz
\begin{subequations}
\begin{gather}
u(x,t)=V(\omega)\exp\left(\alpha_1 x\right)\\[0.3cm]
\omega(x,t)=t.
\end{gather}
\end{subequations}
The reduced ODE for equation (\ref{Eq-Case1.1})
then takes the form 
\begin{gather}
V_\omega=-i\sqrt{2}\lambda V 
\end{gather}
with the general solution
\begin{gather}
V(\omega)=C_1 \exp\left(-i\sqrt{2}\lambda t\right)
\end{gather}
where $C_1$ is a constant of integration. A special solution of (\ref{Eq-Case1.1}) is then
\begin{gather}
u(x,t)=C_1 \exp\left(\alpha_1 x-i\sqrt{2}\lambda t\right)\quad \mbox{for}\ \alpha_1=1.
\end{gather}

\strut\hfill

\noindent
{\bf 4.3 The subalgebra spanned by $\{Z_1+\alpha_1 Z_7\}$} leads to the symmetry Ansatz
\begin{subequations}
\begin{gather}
u(x,t)=V(\omega)\\[0.3cm]
\omega(x,t)=t-\alpha_1 x.
\end{gather}
\end{subequations}
The reduced ODE  for equation (\ref{Eq-Case1.1})
then takes the form
\begin{gather}
\sqrt{2}\lambda V_\omega=-\left(
\vphantom{\frac{da}{db}}
2V_{\omega}V_{3\omega}-3V_{\omega\omega}^2\right)^{1/2}
\end{gather}
with the general solution
\begin{gather}
V(\omega)=C_2\left[C_1-4\lambda \tan\left(
\frac{\lambda}{\sqrt{2}}(\omega+C3)\right)\right]
\end{gather}
where $C_1,\ C_2$ and $C_3$ are constants of integration with $C_2>0$. A special solution of (\ref{Eq-Case1.1}) is then
\begin{gather}
u(x,t)=C_2\left[C_1-4\lambda \tan\left(
\frac{\lambda}{\sqrt{2}}(t-\alpha_1 x+C3)\right)\right],
\end{gather}
whereby $\alpha_1=1$ for $\lambda<0$ and $\alpha_1=-1$ for $\lambda>0$.

\strut\hfill

\noindent
{\bf 4.4 The subalgebra spanned by $\{Z_2+a_1 Z_7\}$} leads to the symmetry Ansatz
\begin{subequations}
\begin{gather}
u(x,t)=V(\omega)\\[0.3cm]
\omega(x,t)=t-a_1 \ln(x),\quad x>0.
\end{gather}
\end{subequations}
The reduced ODE for equation (\ref{Eq-Case1.1})
then takes the form
\begin{gather}
\sqrt{2}a_1\lambda V_\omega=-\left(
\vphantom{\frac{da}{db}}
2a_1^2V_\omega V_{3\omega}  
-3a_1^2V_{\omega\omega}^2   
+V_\omega^2\right)^{1/2}
\end{gather}
with the general solution
\begin{gather}
V(\omega)=C_1\left[
2a_1\sqrt{2a_1^2\lambda^2-1}\, \tan \left(
\frac{\sqrt{2a_1^2\lambda^2-1}(\omega+C_3)}{2a_1}\right)-C_2\right]
\end{gather}
where $C_1,\ C_2$ and $C_3$ are constants of integration with $C_1\neq 0$. A special solution of (\ref{Eq-Case1.1}) is then
\begin{gather}
u(x,t)=C_1\left[
2a_1\sqrt{2a_1^2\lambda^2-1}\, \tan \left(
\frac{\sqrt{2a_1^2\lambda^2-1}\left(t-a_1\ln(x)+C_3\right)}{2a_1}\right)-C_2\right]
\end{gather}
whereby sgn$(x)|a_1\lambda|+a_1\lambda=0$.

\strut\hfill

\noindent
{\bf 4.5 The subalgebra spanned by $\{Z_4+\alpha_1 Z_7\}$} leads to the symmetry Ansatz
\begin{subequations}
\begin{gather}
u(x,t)=V(\omega)+\frac{t}{\alpha_1}\\[0.3cm]
\omega(x,t)=x.
\end{gather}
\end{subequations}
The reduced ODE for equation (\ref{Eq-Case1.1})
then takes the form
\begin{gather}
\sqrt{2}\alpha_1\lambda V_\omega^2=\left(
\vphantom{\frac{da}{db}}
2V_\omega V_{3\omega}-3V_{\omega\omega}^2\right)^{1/2}
\end{gather}
with the general solution
\begin{gather}
V(\omega)=-\frac{\sqrt{2}}{\alpha_1\lambda}\, \mbox{arctanh}\left(
\frac{C_1 (\omega+C_2)}{\alpha_1\lambda} \right)+C_3
\end{gather}
where $C_1,\ C_2$ and $C_3$ are constants of integration with $C_1\neq 0$. A special solution of (\ref{Eq-Case1.1}) is then
\begin{gather}
u(x,t)=\frac{t}{\alpha_1}-\frac{\sqrt{2}}{\alpha_1\lambda}\, \mbox{arctanh}\left(
\frac{C_1 (x+C_2)}{\alpha_1\lambda} \right)+C_3.
\end{gather}

\strut\hfill

\noindent
{\bf 4.6 The subalgebra spanned by $\{Z_5+a_1 Z_7\}$} leads to the symmetry Ansatz
\begin{subequations}
\begin{gather}
u(x,t)=\exp\left(\frac{t}{a_1}\right)V(\omega)\\[0.3cm]
\omega(x,t)=x.
\end{gather}
\end{subequations}
The reduced ODE for equation (\ref{Eq-Case1.1})
then takes the form 
\begin{gather}
\label{ODE15-EqV}
\sqrt{2}\alpha_1\lambda V_\omega^2=V\left(
\vphantom{\frac{da}{db}}
2V_\omega V_{3\omega}-3V_{\omega\omega}^2\right)^{1/2}.
\end{gather}
Since equation (\ref{ODE15-EqV}) does not depend explicitly on $\omega$ it can be reduced further by introducing a new dependent variable $H$, namely
\begin{gather}
\label{Ansatz-15}
H(V)=\frac{dV}{d\omega}.
\end{gather}
This leads to the 2nd-order equation
\begin{gather}
\sqrt{2}a_1\lambda H^2=V\left(
\vphantom{\frac{da}{db}}
2H^3 H_{VV}-H^2 H_V^2\right)^{1/2},
\end{gather}
 with the general solution
\begin{gather}
H(V)=
C_1V+\frac{C_1^2}{4C_2}
V^{1+\sqrt{1-2a_1^2\lambda^2}}
+C_2V^{1-\sqrt{1-2a_1^2\lambda^2}}.
\end{gather}
Here $C_1$ and $C_2\neq 0$ are constants of integration. Using this solution $H(V)$ and integrating (\ref{Ansatz-15}) then leads to a special solution of (\ref{Eq-Case1.1}).

\strut\hfill

\noindent
{\bf 4.7 The subalgebra spanned by $\{Z_1+\alpha_1 Z_3+\alpha_2 Z_4\}$} leads to the symmetry Ansatz
\begin{subequations}
\begin{gather}
u(x,t)=V(\omega)+\frac{\alpha_2}{\sqrt{\alpha_1}}\arctan(\sqrt{\alpha_1} x)
\\[0.3cm]
\omega(x,t)=t.
\end{gather}
\end{subequations}
The reduced ODE for equation (\ref{Eq-Case1.1})
then takes the form
\begin{gather}
V_\omega=-\frac{i\lambda\alpha_2}{\sqrt{2\alpha_1}},\quad i^2=-1,
\end{gather}
with the general solution
\begin{gather}
V(\omega)=-\frac{i\lambda\alpha_2}{\sqrt{2\alpha_1}}\, \omega+C_1
\end{gather}
where $C_1$ is a constant of integration. A special complex solution of (\ref{Eq-Case1.1}) is then
\begin{gather}
u(x,t)=-\frac{i\lambda\alpha_2}{\sqrt{2\alpha_1}}\, t+\frac{\alpha_2}{\sqrt{\alpha_1}}\arctan(\sqrt{\alpha_1} x)+C_1.
\end{gather}

\strut\hfill

\noindent
{\bf 4.8 The subalgebra spanned by $\{Z_1+\alpha_1 Z_4+\alpha_2 Z_6\}$} leads to the symmetry Ansatz
\begin{subequations}
\begin{gather}
u(x,t)=\frac{\sqrt{\alpha_1\alpha_2}}{\alpha_2}\tan\left[
\vphantom{\frac{da}{db}}
\sqrt{\alpha_1\alpha_2}\left(
V(\omega)+x\right)\right]
\\[0.3cm]
\omega(x,t)=t.
\end{gather}
\end{subequations}
The reduced ODE for equation (\ref{Eq-Case1.1})
then takes the form
\begin{gather}
V_\omega=\frac{\lambda}{\sqrt{2}}\frac{1}{\sqrt{\alpha_1\alpha_2}}
\end{gather}
with the general solution
\begin{gather}
V(\omega)=\frac{\lambda}{\sqrt{2}}\frac{1}{\sqrt{\alpha_1\alpha_2}} \omega+C_1
\end{gather}
where $C_1$ is a constant of integration. A special solution of (\ref{Eq-Case1.1}) is then
\begin{gather}
u(x,t)=\frac{\sqrt{\alpha_1\alpha_2}}{\alpha_2}\tan\left[
\vphantom{\frac{da}{db}}
\sqrt{\alpha_1\alpha_2}\left(
 \frac{\lambda}{\sqrt{2}}\frac{t}{\sqrt{\alpha_1\alpha_2}} +x+C_1\right)\right]
\end{gather}

\strut\hfill

\noindent
{\bf 4.9 The subalgebra spanned by $\{Z_1+\alpha_1 Z_3+a_1 Z_7\}$} leads to the symmetry Ansatz
\begin{subequations}
\begin{gather}
u(x,t)=V(\omega)
\\[0.3cm]
\omega(x,t)=t-\frac{a_1}{\sqrt{\alpha_1}}\left[
\vphantom{\frac{da}{db}}
\sqrt{\alpha_1}\,t-\frac{a_1}{\sqrt{\alpha_1}}\arctan\left(\sqrt{\alpha_1}\,x\right)\right].
\end{gather}
\end{subequations}
The reduced ODE for equation (\ref{Eq-Case1.1})
then takes the form 
\begin{gather}
\label{ODE-18-EqV}
\sqrt{2}a_1\lambda V_\omega =-\left(
\vphantom{\frac{da}{db}}
2a_1^2 V_\omega V_{3\omega}
-3a_1^2 V_{\omega\omega}^2
-4\alpha_1 V_\omega^2\right)^{1/2}.
\end{gather}
Since equation (\ref{ODE-18-EqV}) does not depend explicitly on $V$ it can be reduced further by introducing a new dependent variable $P$, namely
\begin{gather}
\label{Ansatz-18-P}
P(\omega)=\frac{dV}{d\omega}.
\end{gather}
This leads to the 2nd-order equation
\begin{gather}
\label{ODE-18-EqP}
\sqrt{2}a_1\lambda P =-\left(
\vphantom{\frac{da}{db}}
2a_1^2 P P_{\omega\omega}
-3a_1^2 P_\omega^2
-4\alpha_1 P^2\right)^{1/2}.
\end{gather}
Since equation (\ref{ODE-18-EqP}) does not depend explicitly on $\omega$ it can be reduced further by introducing a new dependent variable $H$, namely
\begin{gather}
\label{Ansatz-18-H}
H(P)=\frac{dP}{d\omega}.
\end{gather}
This leads to the 1st-order equation
\begin{gather}
\label{ODE-18-EqH}
\sqrt{2}a_1\lambda P =-\left(
\vphantom{\frac{da}{db}}
2a_1^2 P H\frac{dH}{dP}
-3a_1^2 H^2
-4\alpha_1 P^2\right)^{1/2}
\end{gather}
with the general solution
\begin{gather}
H(P)=\pm \frac{P}{a_1}
\left(
\vphantom{\frac{da}{db}}
a_1^2C_1P-2a_1^2\lambda^2-4\alpha_1\right)^{1/2},
\end{gather}
where $C_1$ is a constant of integration. Using this solution $H(P)$ and integrating (\ref{Ansatz-18-H}) and 
(\ref{Ansatz-18-P}) then leads to a special solution of (\ref{Eq-Case1.1}).

\strut\hfill

\noindent
{\bf 4.10 The subalgebra spanned by $\{Z_1+\alpha_1 Z_4+a_1 Z_7\}$} leads to the symmetry Ansatz
\begin{subequations}
\begin{gather}
\label{Ansatz-case19-a}
u(x,t)=V(\omega)+\alpha_1 x
\\[0.3cm]
\label{Ansatz-case19-b}
\omega(x,t)=t-a_1 x.
\end{gather}
\end{subequations}
The reduced ODE for equation (\ref{Eq-Case1.1})
then takes the form
\begin{gather}
\label{Reduce-1-case19}
-\sqrt{2}\lambda \left(
a_1V_\omega-\alpha_1
\right)^2
=a_1^{3/2}V_\omega
\left(
\vphantom{\frac{da}{db}}
2a_1V_\omega V_{3\omega}
-2\alpha_1V_{3\omega}
-3a_1V_{\omega\omega}^2
\right)^{1/2}.
\end{gather}
Since equation (\ref{Reduce-1-case19}) does not depend explicitly on $V(\omega)$, we can reduce this equation further by introducing a new dependent variable $P$, namely
\begin{gather}
\label{case19-V}
P(\omega)=\frac{dV}{d\omega}.
\end{gather}
This leads to the 2nd-order ODE
\begin{gather}
\label{Reduce-2-case19}
-\sqrt{2}\lambda \left(
a_1P-\alpha_1
\right)^2
=a_1^{3/2}P
\left(
\vphantom{\frac{da}{db}}
2a_1P P_{3\omega}
-2\alpha_1V_{\omega\omega}
-3a_1P_{\omega}^2
\right)^{1/2}.
\end{gather}
Since equation (\ref{Reduce-2-case19}) does not depend explicitly on $\omega$ it can be reduced further by introducing a new dependent variable $H$, namely
\begin{gather}
\label{case 19-P}
H(P)=\frac{dP}{d\omega},
\end{gather}
whereby (\ref{Reduce-2-case19}) reduces the the following 1st-order ODE:
\begin{gather}
\label{Reduce-3-case19}
-\sqrt{2}\lambda \left(
a_1P-\alpha_1
\right)^2=a_1^{3/2}P
\left(
\vphantom{\frac{da}{db}}
-2a_1P H\frac{dH}{dP}
-2\alpha_1H\frac{dH}{dP}
-3a_1H^2
\right)^{1/2}.
\end{gather}
The general solution of equation (\ref{Reduce-3-case19}) is
\begin{gather}
H(P)=\pm \frac{a_1P-\alpha_1}{a_1^2P^2}\left[
\vphantom{\frac{da}{db}}
a_1P\left(a_1^4C_1P^2-a_1^3\alpha_1C_1P
-2a_1\lambda^2P
+2\alpha_1\lambda^2\right)\right]^{1/2}
\end{gather}
where $C_1$ is a constant of integration. This solutions can now be used to find $P$ by integrating (\ref{case 19-P}),
and $V(\omega)$ by integrating (\ref{case19-V}). Finally a solution for (\ref{Eq-Case1.1}) then follows from the symmetry Ansatz (\ref{Ansatz-case19-a})--(\ref{Ansatz-case19-b})

\strut\hfill

\noindent
{\bf 4.11 The subalgebra spanned by $\{Z_2+\alpha_1 Z_4+a_1 Z_7\}$} leads to the symmetry Ansatz
\begin{subequations}
\begin{gather}
\label{Ansatz-case20-a}
u(x,t)=V(\omega)+\alpha_1 \ln(x)
\\[0.3cm]
\label{Ansatz-case20-b}
\omega(x,t)=t-a_1 \ln(x),\quad x>0.
\end{gather}
\end{subequations}
The reduced ODE for equation (\ref{Eq-Case1.1})
then takes the form
\begin{gather}
%\label{Reduce-1-case20}
-\sqrt{2}\lambda \left(
a_1V_\omega-\alpha_1
\right)^2
=V_\omega
\left(
\vphantom{\frac{da}{db}}
2a_1^4V_\omega V_{3\omega}
-2a_1^3\alpha_1V_{3\omega}
-3a_1^4V_{\omega\omega}^2
+a_1^2V_\omega^2
\right.\nn\\[0.3cm]
\label{Reduce-1-case20}
\qquad 
\left.
\vphantom{\frac{da}{db}}
-2a_1\alpha_1 V_\omega
+1
\right)^{1/2}.
\end{gather}
Since equation (\ref{Reduce-1-case20}) does not depend explicitly on $V(\omega)$, we can reduce this equation further by introducing a new dependent variable $P$, namely
\begin{gather}
\label{case20-V}
P(\omega)=\frac{dV}{d\omega}.
\end{gather}
This leads to the 2nd-order ODE
\begin{gather}
%\label{Reduce-2-case20}
-\sqrt{2}\lambda \left(
a_1P-\alpha_1
\right)^2
=P
\left(
\vphantom{\frac{da}{db}}
2a_1^4P P_{\omega\omega}
-2a_1^3\alpha_1P_{\omega\omega}
-3a_1^4P_{\omega}^2
+a_1^2P^2
\right.\nn\\[0.3cm]
\label{Reduce-2-case20}
\qquad 
\left.
\vphantom{\frac{da}{db}}
-2a_1\alpha_1 P
+1
\right)^{1/2}.
\end{gather}
Since equation (\ref{Reduce-2-case20}) does not depend explicitly on $\omega$ it can be reduced further by introducing a new dependent variable $H$, namely
\begin{gather}
\label{case 20-P}
H(P)=\frac{dP}{d\omega},
\end{gather}
whereby (\ref{Reduce-2-case20}) reduces the the following 1st-order ODE:
\begin{gather}
%\label{Reduce-3-case20}
-\sqrt{2}\lambda \left(
a_1P-\alpha_1
\right)^2
=P
\left(
\vphantom{\frac{da}{db}}
2a_1^4PH\frac{dH}{dP}
-2a_1^3\alpha_1H\frac{dH}{dP}
-3a_1^4H^2
+a_1^2P^2
\right.\nn\\[0.3cm]
\label{Reduce-3-case20}
\qquad 
\left.
\vphantom{\frac{da}{db}}
-2a_1\alpha_1 P
+1
\right)^{1/2}.
\end{gather}
The general solution of equation (\ref{Reduce-3-case19}) is
\begin{gather}
H(P)=\pm \frac{a_1P-\alpha_1}{a_1^2P^2}\left[
\vphantom{\frac{da}{db}}
P\left(a_1^5C_1P^2
-a_1^4\alpha_1C_1P
-2a_1^2\lambda^2P
+2a_1\alpha_1\lambda^2+P\right)\right]^{1/2}
\end{gather}
where $C_1$ is an arbitrary constant. This solutions can now be used to find $P$ by integrating (\ref{case 20-P}),
and $V(\omega)$ by integrating (\ref{case20-V}). Finally a solution for (\ref{Eq-Case1.1}) then follows from the symmetry Ansatz (\ref{Ansatz-case20-a})--(\ref{Ansatz-case20-b}).

\strut\hfill

\noindent
{\bf 4.12 The subalgebra spanned by $\{Z_1+\alpha_1 Z_5+a_1 Z_7\}$} leads to the symmetry Ansatz
\begin{subequations}
\begin{gather}
\label{Ansatz-case21-a}
u(x,t)=V(\omega) \exp(\alpha_1 x)
\\[0.3cm]
\label{Ansatz-case21-b}
\omega(x,t)=t-a_1 x.
\end{gather}
\end{subequations}
The reduced ODE for equation (\ref{Eq-Case1.1})
then takes the form
\begin{gather}
%\label{Reduce-1-case21}
\sqrt{2}\lambda \left(
a_1V_\omega-\alpha_1 V
\right)^2
=V_\omega
\left[
\vphantom{\frac{da}{db}}
\left(
2a_1^4V_{3\omega}+6a_1^3\alpha_1 V_{\omega\omega}+4a_1\alpha_1 V\right)V_\omega
-2a_1^3\alpha_1V V_{3\omega}
\right.
\nn\\[0.3cm]
\label{Reduce-1-case21}
\qquad
\left.
\vphantom{\frac{da}{db}}
-6a_1^2V_\omega^2
-3a_1^4V_{\omega\omega}^2
-V^2\right]^{1/2}
\end{gather}
Since equation (\ref{Reduce-1-case21}) does not depend explicitly on $\omega$, we can reduce this equation further by introducing a new dependent variable $H$, namely
\begin{gather}
\label{case21-V}
H(V)=\frac{dV}{d\omega}.
\end{gather}
This leads to the 2nd-order ODE
\begin{gather}
%\label{Reduce-2-case21}
\sqrt{2}\lambda \left(
a_1H-\alpha_1 V
\right)^2
=H
\left[
\vphantom{\frac{da}{db}}
2a_1^3H^2(a_1H-\alpha_1 V)\frac{d^2H}{dV^2}
-a_1^3H(a_1 H+2\alpha_1V)
\left(\frac{dH}{dV}\right)^2
\right.\nn\\[0.3cm]
\label{Reduce-2-case21}
\qquad
\left.
\vphantom{\frac{da}{db}}
+6a_1^3\alpha_1H^2\frac{dH}{dV}
-6a_1^2H^2+4a_1\alpha_1V H-V^2\right]^{1/2}.
\end{gather}
Equation (\ref{Reduce-2-case21}) is solvable but the integrals depend on the values of the constants in the equations, so we will not take this any further here.

%%%%
%%%%%

\strut\hfill

\noindent
{\bf 4.13 The subalgebra spanned by $\{Z_2+a_1 Z_5+a_2 Z_7\}$} leads to the symmetry Ansatz
\begin{subequations}
\begin{gather}
\label{Ansatz-case22-a}
u(x,t)=x^{a_1}V(\omega) 
\\[0.3cm]
\label{Ansatz-case22-b}
\omega(x,t)=t-a_1 \ln(x),\quad x>0.
\end{gather}
\end{subequations}
The reduced ODE for equation (\ref{Eq-Case1.1})
then takes the form
\begin{gather}
%\label{Reduce-1-case22}
\sqrt{2}\lambda \left(
a_2V_\omega-a_1 V
\right)^2
=V_\omega
\left[
\vphantom{\frac{da}{db}}
\left(
\vphantom{\frac{da}{db}}
2a_1^4V_{3\omega}
+6a_1a_2^3V_{\omega\omega}^2
+2a_1a_2(2a_1^2-1)V_\omega
\right)V_\omega
\right.\nn\\[0.3cm]
\qquad
\label{Reduce-1-case22}
\left.
\vphantom{\frac{da}{db}}
-a_1a_2^3VV_{3\omega}
-3a_2^4V_{\omega\omega}^2
-a_1^2(a_1^2-1)V^2
\right]^{1/2}.
\end{gather}
Since equation (\ref{Reduce-1-case22}) does not depend explicitly on $\omega$, we can reduce this equation further by introducing a new dependent variable $H$, namely
\begin{gather}
\label{case21-V}
H(V)=\frac{dV}{d\omega}.
\end{gather}
This leads to the 2nd-order ODE
\begin{gather}
%\label{Reduce-2-case22}
\sqrt{2}\lambda \left(
a_2H-a_1 V
\right)^2
=
H\left[
\vphantom{\frac{da}{db}}
2a_2^3H^2(a_2H-a_1V)\frac{d^2H}{dV^2}
-2a_2^3H(a_2H+a_1V)\left(\frac{dH}{dV}\right)^2
\right.\nn\\[0.3cm]
\label{Reduce-2-case22}
\qquad
\left.
\vphantom{\frac{da}{db}}
+6a_1a_2^3H^2\frac{dH}{dV}
+a_2^2(1-6a_1^2)H^2
+2a_2a_1V(2a_1^2-1)H
-a_1^2(a_1^2-1)V^2\right]^{1/2}.
%-a_1^4V^2
%+a_1^2V^2\right]^{1/2}
\end{gather}
Equation (\ref{Reduce-2-case22}) is solvable but the integrals depend on the values of the constants in the equations, so we will not take this any further here.

%%%%
%%%%%

\strut\hfill

\noindent
{\bf 4.14 The subalgebra spanned by $\{Z_4+\alpha_1 Z_6+a_1 Z_7\}$} leads to the symmetry Ansatz
\begin{subequations}
\begin{gather}
\label{Ansatz-case23-a}
u(x,t)=\frac{1}{\sqrt{\alpha_1}}\tan\left[
\frac{\sqrt{\alpha_1}}{a_1}(t+V(\omega))\right]
\\[0.3cm]
\label{Ansatz-case23-b}
\omega(x,t)=x.
\end{gather}
\end{subequations}
The reduced ODE for equation (\ref{Eq-Case1.1})
then takes the form
\begin{gather}
\label{Reduce-1-case23}
\sqrt{2}\lambda a_1V_\omega^2
=
\left(
\vphantom{\frac{da}{db}}
2a_1^2V_\omega V_{3\omega}
-3a_1^2V_{\omega\omega}^2
+4\alpha_1V_\omega^4
\right)^{1/2}.
\end{gather}
Since equation (\ref{Reduce-1-case23}) does not depend explicitly on $V(\omega)$, we can reduce this equation further by introducing a new dependent variable $P$, namely
\begin{gather}
\label{case23-V}
P(\omega)=\frac{dV}{d\omega}.
\end{gather}
This leads to the 2nd-order ODE
\begin{gather}
\label{Reduce-2-case23}
\sqrt{2}\lambda a_1P^2
=
\left(
\vphantom{\frac{da}{db}}
2a_1^2P P_{\omega\omega}
-3a_1^2P_\omega^2
+4\alpha_1P^4
\right)^{1/2}.
\end{gather}
Since equation (\ref{Reduce-2-case23}) does not depend explicitly on $\omega$ it can be reduced further by introducing a new dependent variable $H$, namely
\begin{gather}
\label{case 23-P}
H(P)=\frac{dP}{d\omega},
\end{gather}
whereby (\ref{Reduce-2-case23}) reduces the the following 1st-order ODE:
\begin{gather}
\label{Reduce-3-case23}
\sqrt{2}\lambda a_1P^2
=\left[
\vphantom{\frac{da}{db}}
2a_1^2PH\frac{dH}{dP}
-3a_1^2H^2
+4\alpha_1 P^4\right]^{1/2}
\end{gather}
The general solution of equation (\ref{Reduce-3-case23}) is
\begin{gather}
H(P)=\pm
\frac{P}{a_1}\left[
\vphantom{\frac{da}{db}}
2a_1^2\lambda^2P^2-4\alpha_1 P^2+C_1a_1^2 P\right]^{1/2},
\end{gather}
where $C_1$ is a constant of integration. This solutions can now be used to find $P$ by integrating (\ref{case 23-P}),
and $V(\omega)$ by integrating (\ref{case23-V}). Finally a solution for (\ref{Eq-Case1.1}) then follows from the symmetry Ansatz (\ref{Ansatz-case23-a})--(\ref{Ansatz-case23-b}).

\strut\hfill

\noindent
{\bf 4.15 The subalgebra spanned by $\{Z_1+\alpha_1 Z_3+\alpha_2 Z_4+a_1 Z_6\}$} leads to the symmetry Ansatz
\begin{subequations}
\begin{gather}
\label{Ansatz-case24-a}
u(x,t)=\frac{\sqrt{a_1\alpha_2}}{a_1}
\tan \left[
\vphantom{\frac{da}{db}}
\frac{\sqrt{a_1\alpha_2}}{\sqrt{\alpha_1}}
\left(
\vphantom{\frac{da}{db}}
\sqrt{\alpha_1}\,V(\omega)
+
\arctan(\sqrt{\alpha_1}x)\right)
\right]
\\[0.3cm]
\omega(x,t)=t.
\end{gather}
\end{subequations}
The reduced ODE for equation (\ref{Eq-Case1.1})
then takes the form
\begin{gather}
\label{Reduce-1-case24}
V_\omega=\frac{\lambda}{
\sqrt{2}(a_1\alpha_2-\alpha_1)^{1/2}},\quad a_1\alpha_2-\alpha_1\neq 0
\end{gather}
with the general solution
\begin{gather}
V(\omega)=\frac{\lambda \omega}{
\sqrt{2}(a_1\alpha_2-\alpha_1)^{1/2}}+C_1,
\end{gather}
where $C_1$ is a constant of integration. This leads to the following solution for (\ref{Eq-Case1.1}):
\begin{gather}
u(x,t)=\frac{\sqrt{a_1\alpha_2}}{a_1}
\tan \left[
\vphantom{\frac{da}{db}}
\frac{\sqrt{a_1\alpha_2}}{\sqrt{\alpha_1}}
\left(
\frac{ \sqrt{\alpha_1}\lambda t}{
\sqrt{2}(a_1\alpha_2-\alpha_1)^{1/2}}
+
\arctan(\sqrt{\alpha_1}x)
+C_1
\right)
\right].
\end{gather}

\strut\hfill

\noindent
{\bf 4.16 The subalgebra spanned by $\{Z_1+\alpha_1 Z_3+\alpha_2 Z_4+ a_1Z_7\}$} leads to the symmetry Ansatz
\begin{subequations}
\begin{gather}
\label{Ansatz-case25-a}
u(x,t)=V(\omega)+\frac{\alpha_2}{\sqrt{\alpha_1}}\arctan (\sqrt{\alpha_1} x)
\\[0.3cm]
\label{Ansatz-case25-b}
\omega(x,t)=t-\frac{a_1}{\sqrt{\alpha_1}}\arctan (\sqrt{\alpha_1} x)
\end{gather}
\end{subequations}
The reduced ODE for equation (\ref{Eq-Case1.1})
then takes the form
\begin{gather}
%\label{Reduce-1-case25}
\sqrt{2}\lambda (a_1V_\omega -\alpha_2)^2
=-V_\omega\left[
\vphantom{\frac{da}{db}}
2a_1^4V_\omega V_{3\omega}
-2a_1^3\alpha_2V_{3\omega}
-3a_1^4V_{\omega\omega}^2
-4a_1^2\alpha_1V_\omega^2\right.
\nn\\[0.3cm]
\label{Reduce-1-case25}
\qquad
\left.
\vphantom{\frac{da}{db}}
+8a_1\alpha_1\alpha_2V_\omega
-4\alpha_1\alpha_2^2
\right]^{1/2}.
\end{gather}
Since equation (\ref{Reduce-1-case25}) does not depend explicitly on $V(\omega)$, we can reduce this equation further by introducing a new dependent variable $P$, namely
\begin{gather}
\label{case25-V}
P(\omega)=\frac{dV}{d\omega}.
\end{gather}
This leads to the 2nd-order ODE
\begin{gather}
%\label{Reduce-2-case25}
\sqrt{2}\lambda (a_1P -\alpha_2)^2
=-P\left[
\vphantom{\frac{da}{db}}
2a_1^4P P_{\omega\omega}
-2a_1^3\alpha_2P_{\omega\omega}
-3a_1^4P_{\omega}^2
-4a_1^2\alpha_1P^2\right.
\nn\\[0.3cm]
\label{Reduce-2-case25}
\qquad
\left.
\vphantom{\frac{da}{db}}
+8a_1\alpha_1\alpha_2P
-4\alpha_1\alpha_2^2
\right]^{1/2}.
\end{gather}
Since equation (\ref{Reduce-2-case25}) does not depend explicitly on $\omega$ it can be reduced further by introducing a new dependent variable $H$, namely
\begin{gather}
\label{case 25-P}
H(P)=\frac{dP}{d\omega},
\end{gather}
whereby (\ref{Reduce-2-case25}) reduces the the following 1st-order ODE:
\begin{gather}
%\label{Reduce-3-case25}
\sqrt{2}\lambda (a_1P -\alpha_2)^2
=-P\left[
\vphantom{\frac{da}{db}}
2a_1^4PH\frac{dH}{dP}
-2a_1^3\alpha_3H\frac{dH}{dP}
-3a_1^4H^2
-4a_1^2\alpha_1P^2\right.\nn\\[0.3cm]
\label{Reduce-3-case25}
\qquad
\left.
\vphantom{\frac{da}{db}}
+8a_1\alpha_1\alpha_2P
-4\alpha_1\alpha_2^2\right]^{1/2}
\end{gather}
The general solution of equation (\ref{Reduce-3-case25}) is
\begin{gather}
H(P)=\pm\frac{a_1P-\alpha_2}{a_1^2P}
\left[
\vphantom{\frac{da}{db}}
P\left(
a_1^5C_1P^2
-a_1^4\alpha_2C_1P
-2a_1^2\lambda^2P
-4\alpha_1 P
+2a_1\alpha_2\lambda^2\right)\right]^{1/2}
\end{gather}
where $C_1$ is a constant of integration. This solutions can now be used to find $P$ by integrating (\ref{case 25-P}),
and $V(\omega)$ by integrating (\ref{case25-V}). Finally a solution for (\ref{Eq-Case1.1}) then follows from the symmetry Ansatz (\ref{Ansatz-case25-a})--(\ref{Ansatz-case25-b}).

\strut\hfill

\noindent
{\bf 4.17. The subalgebra spanned by $\{Z_1+\alpha_1 Z_4+\alpha_2 Z_6+ a_1 Z_7\}$} leads to the symmetry Ansatz
\begin{subequations}
\begin{gather}
\label{Ansatz-case26-a}
u(x,t)=\frac{\sqrt{\alpha_1\alpha_2}}{\alpha_2}\tan\left[
\vphantom{\frac{da}{db}}
\sqrt{\alpha_1\alpha_2}\left( V(\omega)+x\right)\right]
\\[0.3cm]
\label{Ansatz-case26-b}
\omega(x,t)=t-a_1 x.
\end{gather}
\end{subequations}
The reduced ODE for equation (\ref{Eq-Case1.1})
then takes the form
\begin{gather}
%\label{Reduce-1-case26}
\sqrt{2}\lambda (a_1V_\omega -1)^2
=V_\omega\left[
\vphantom{\frac{da}{db}}
2a_1^4V_\omega V_{3\omega}
-2a_1^3V_{3\omega}
-3a_1^4V_{\omega\omega}^2
+4a_1^4\alpha_1\alpha_2V_\omega^4
\right.
\nn\\[0.3cm]
\label{Reduce-1-case26}
\qquad
\left.
\vphantom{\frac{da}{db}}
-16a_1^3\alpha_1\alpha_2V_\omega^3
+24a_1^2\alpha_1\alpha_2V_\omega^2
-16a_1\alpha_1\alpha_2V_\omega
+4a_1\alpha_2
\right]^{1/2}.
\end{gather}
Since equation (\ref{Reduce-1-case26}) does not depend explicitly on $V(\omega)$, we can reduce this equation further by introducing a new dependent variable $P$, namely
\begin{gather}
\label{case26-V}
P(\omega)=\frac{dV}{d\omega}.
\end{gather}
This leads to the 2nd-order ODE
\begin{gather}
%\label{Reduce-2-case25}
\sqrt{2}\lambda (a_1P -1)^2
=V_\omega\left[
\vphantom{\frac{da}{db}}
2a_1^4P P_{\omega\omega}
-2a_1^3P_{\omega\omega}
-3a_1^4P_{\omega}^2
+4a_1^4\alpha_1\alpha_2P^4
\right.
\nn\\[0.3cm]
\label{Reduce-2-case26}
\qquad
\left.
\vphantom{\frac{da}{db}}
-16a_1^3\alpha_1\alpha_2P^3
+24a_1^2\alpha_1\alpha_2P^2
-16a_1\alpha_1\alpha_2P
+4a_1\alpha_2
\right]^{1/2}.
\end{gather}
Since equation (\ref{Reduce-2-case26}) does not depend explicitly on $\omega$ it can be reduced further by introducing a new dependent variable $H$, namely
\begin{gather}
\label{case 26-P}
H(P)=\frac{dP}{d\omega},
\end{gather}
whereby (\ref{Reduce-2-case26}) reduces the the following 1st-order ODE:
\begin{gather}
%\label{Reduce-3-case26}
\sqrt{2}\lambda (a_1P -1)^2
=-P\left[
\vphantom{\frac{da}{db}}
2a_1^4PH\frac{dH}{dP}
-2a_1^3H\frac{dH}{dP}
-3a_1^4H^2
+4a_1^4\alpha_1\alpha_2 P^4
\right.\nn\\[0.3cm]
\label{Reduce-3-case26}
\qquad
\left.
\vphantom{\frac{da}{db}}
-16a_1^3\alpha_1\alpha_2 P^3
+24a_1^2\alpha_1\alpha_2P^2
-16a_1\alpha_1\alpha_2 P
+4\alpha_1\alpha_2
\right]^{1/2}
\end{gather}
The general solution of equation (\ref{Reduce-3-case26}) is
\begin{gather}
H(P)=\pm\frac{a_1P-1}{a_1^2P}
\left[
\vphantom{\frac{da}{db}}
a_1P\left(
-4a_1\alpha_1\alpha_2 P^3
+a_1^4C_1 P^2
+4\alpha_1\alpha_2 P^2
-a_1^3C_1 P
\right.\right.
\nn\\[0.3cm]
\qquad
\left.
\vphantom{\frac{da}{db}}
\left.
-2a_1\lambda^2 P
+2\lambda^2
\right)\right]^{1/2},
\end{gather}
where $C_1$ is a constant of integration. This solutions can now be used to find $P$ by integrating (\ref{case 26-P}),
and $V(\omega)$ by integrating (\ref{case26-V}). Finally a solution for (\ref{Eq-Case1.1}) then follows from the symmetry Ansatz (\ref{Ansatz-case26-a})--(\ref{Ansatz-case26-b}).

\strut\hfill

\noindent
{\bf 4.18 The subalgebra spanned by $\{Z_1+\alpha_1 Z_3+\alpha_2 Z_4+ a_1 Z_6+a_2 Z_7\}$} leads to the symmetry Ansatz
\begin{subequations}
\begin{gather}
\label{Ansatz-case27-a}
u(x,t)=\frac{\sqrt{a_1\alpha_2}}{a_1}\tan\left[
\vphantom{\frac{da}{db}}
\frac{\sqrt{a_1\alpha_2}}{\sqrt{\alpha_1}}
\left(
\vphantom{\frac{da}{db}}
\sqrt{\alpha_1}\,\,V(\omega)+
\arctan (\sqrt{\alpha_1}\, x)\right)\right]
\\[0.3cm]
\label{Ansatz-case27-b}
\omega(x,t)=t-\frac{a_2}{\sqrt{\alpha_1}} \arctan(\sqrt{\alpha_1}\, x).
\end{gather}
\end{subequations}
The reduced ODE for equation (\ref{Eq-Case1.1})
then takes the form
\begin{gather}
%\label{Reduce-1-case27}
\sqrt{2}\lambda (a_2V_\omega -1)^2
=V_\omega\left[
\vphantom{\frac{da}{db}}
\left(
\vphantom{\frac{da}{db}}
2a_2^4 V_{3\omega}-16a_2\left(a_1\alpha_2-\frac{\alpha_1}{2}\right)
\right)V_\omega
-2a_2^3V_{3\omega}
-3a_2^4V_{\omega\omega}^2
\right.
\nn\\[0.3cm]
\label{Reduce-1-case27}
\qquad
\left.
\vphantom{\frac{da}{db}}
+4a_1a_2^4\alpha_2 V_\omega^4
-16a_1a_2^3\alpha_2 V_\omega^3
+24a_2^2\left(a_1\alpha_2-\frac{\alpha_1}{6}\right)V_\omega^2
+4a_1\alpha_2-4\alpha_1
\right]^{1/2}.
\end{gather}
Since equation (\ref{Reduce-1-case27}) does not depend explicitly on $V(\omega)$, we can reduce this equation further by introducing a new dependent variable $P$, namely
\begin{gather}
\label{case27-V}
P(\omega)=\frac{dV}{d\omega}.
\end{gather}
This leads to the 2nd-order ODE
\begin{gather}
%\label{Reduce-2-case27}
\sqrt{2}\lambda (a_2P -1)^2
= P \left[
\vphantom{\frac{da}{db}}
\left(
\vphantom{\frac{da}{db}}
2a_2^4 P_{\omega\omega}-16a_2\left(a_1\alpha_2-\frac{\alpha_1}{2}\right)
\right)P
-2a_2^3P_{\omega\omega}
-3a_2^4P_{\omega}^2
\right.
\nn\\[0.3cm]
\label{Reduce-2-case27}
\qquad
\left.
\vphantom{\frac{da}{db}}
+4a_1a_2^4\alpha_2 P^4
-16a_1a_2^3\alpha_2 P^3
+24a_2^2\left(a_1\alpha_2-\frac{\alpha_1}{6}\right)P^2
+4a_1\alpha_2-4\alpha_1
\right]^{1/2}.
\end{gather}
Since equation (\ref{Reduce-2-case27}) does not depend explicitly on $\omega$ it can be reduced further by introducing a new dependent variable $H$, namely
\begin{gather}
\label{case 27-P}
H(P)=\frac{dP}{d\omega},
\end{gather}
whereby (\ref{Reduce-2-case27}) reduces the the following 1st-order ODE:
\begin{gather}
%\label{Reduce-3-case27}
\sqrt{2}\lambda (a_2P -1)^2
=P\left[
\vphantom{\frac{da}{db}}
2a_2^3(a_2P-1)H\frac{dH}{dP}
-3a_2^4H^2\right.\nn\\[0.3cm]
\qquad
\label{Reduce-3-case27}
\left.
+4(a_2P-1)^2
(a_1a_2^2\alpha_2 P^2
-2a_1a_2\alpha_2 P
+a_1\alpha_2-\alpha_1)
\right]^{1/2}
\end{gather}
The general solution of equation (\ref{Reduce-3-case27}) is
\begin{gather}
H(P)=\pm\frac{a_2P-1}{a_2^2P}
\left[
\vphantom{\frac{da}{db}}
P\left(
-4a_1a_2^2\alpha_2 P^3
+a_2^5 C_1P^2
+4a_1a_2 \alpha_2P^2
\right.\right.
\nn\\[0.3cm]
\qquad
\left.
\vphantom{\frac{da}{db}}
\left.
-a_2^2C_1P
-2a_2^2\lambda^2 P
-4\alpha_1 P
+2a_2\lambda^2
\right)\right]^{1/2},
\end{gather}
where $C_1$ is a constant of integration. This solutions can now be used to find $P$ by integrating (\ref{case 27-P}),
and $V(\omega)$ by integrating (\ref{case27-V}). Finally a solution for (\ref{Eq-Case1.1}) then follows from the symmetry Ansatz (\ref{Ansatz-case27-a})--(\ref{Ansatz-case27-b}).

\section{The fully-nonlinear equation (\ref{Eq-Case1.1}) and the Schwarzian KdV}
We recall \cite{P-E-E-2004} that the Schwarzian KdV 
\begin{gather}
\label{SKdV-V}
V_T=V_XS[V]
\end{gather}
and the equation
\begin{gather}
\label{3rd-order-Z1Z2Z3}
v_t=\frac{S[v]}{v_x^2},
\end{gather}
are related by the standard hodograph transformation $X=v(x,t)$, $V(X,T)=x$ with $T=t$. We remind 
\cite{Weiss-1983}
the reader 
the that the Schwarzian KdV is obtained from the KdV equation
\begin{gather}
S_T=S_{3X}+3SS_X,
\end{gather}
where $S[V]$ is of course the Schwarzian derivative in terms of $V$
\begin{gather*}
S[V]=\frac{V_{3X}}{V_X}-\frac{3}{2}\frac{V_{XX}^2}{V_X^2}.
\end{gather*}
 Note further that (\ref{SKdV-V}) admits  the Lie symmetry algebra $\{Z_4,Z_5,Z_6\}$ while (\ref{3rd-order-Z1Z2Z3})
 admits $\{Z_1,Z_2,Z_3\}$ (in terms of the appropriate variables of course), so this fact makes this hodograph connection between the two equations rather obvious.
Furthermore, equation (\ref{3rd-order-Z1Z2Z3}) potentialises in
\begin{gather}
\label{HT-invariant-eq}
\tilde v_t=\frac{\tilde v_{3x}}{\tilde v_x^{3/2}}-\frac{3}{2}\frac{\tilde v_{xx}^2}{\tilde v_x^{5/2}}
\end{gather}
with
$\tilde v_x=v_x^2.$
Now, introducing the new dependent variable
$
W(x,t)=\tilde v_x,
$
equation (\ref{HT-invariant-eq}) becomes
\begin{gather}
\label{3rd-order-Eq-W}
%W_t=W^{-3/2}W_{xxx}-\frac{9}{2}W^{-5/2}W_xW_{xx}+\frac{15}{4}W^{-7/2}W_x^3.
W_t=\frac{W_{3x}}{W^{3/2}}-\frac{9}{2}\frac{W_xW_{xx}}{W^{5/2}}+\frac{15}{4}\frac{W_x^3}{W^{7/2}}.
\end{gather}
On the other hand the fully-nonlinear equation (\ref{Eq-Case1.1}), written in terms of the Schwarzian derivative $S$, takes the form
\begin{gather}
\label{3rd-order-Eq-S}
%W_t=W^{-3/2}W_{xxx}-\frac{9}{2}W^{-5/2}W_xW_{xx}+\frac{15}{4}W^{-7/2}W_x^3.
S_t=\frac{S_{3x}}{S^{3/2}}-\frac{9}{2}\frac{S_xS_{xx}}{S^{5/2}}+\frac{15}{4}\frac{S_x^3}{S^{7/2}},
\end{gather}
which is identical to (\ref{3rd-order-Eq-W}), albeit in the variable $S$. 
By combining the above change of variables we obtain the following hodograph transformation
which  provides the mapping between the fully-nonlinear equation (\ref{Eq-Case1.1})  and the Schwarzian-KdV (\ref{SKdV-V}): %\ref{FN-2}
\begin{gather}
\label{HT-2}  
{\cal HT}: 
\left\{
\ba{l}
\displaystyle{
X=\int \sqrt{S[u]}\  dx  
}\\
\\
\displaystyle{ T=t   }\\
\\
\displaystyle{ V(X,T)=x.}\\
\ea
\right.
\end{gather}

\section{Ordinary differential equations invariant under the projective transformation}

We now consider the following projective transformation in the variables $x$ and $u(x)$:
\begin{gather}
\label{Mobius-u-x}  
{\cal M}_2: 
\left\{
\ba{l}
\displaystyle{
u({x})\mapsto v(\bar{x})= \frac{\alpha_1 u({x})+\beta_1}{\alpha_2 u({x})+\beta_2}}\\
\\
\displaystyle{{x}\mapsto \bar {x}=\frac{\gamma_1 x+\delta_1}{\gamma_2 x+\delta_2}  }\\
%\\
%\displaystyle{{t}\mapsto \bar {t}=t+\epsilon}
\ea
\right.
\end{gather}
with the same conditions on the parameters $\alpha_j,\ \beta_j,\ \delta_j$ and $\gamma_j$ as for the transformation (\ref{Mobius-u-x-t}).
The Lie generators corresponding to (\ref{Mobius-u-x}) are
\begin{gather}
\label{Lie-Alg-6dim}
\{Z_1=\pde{\ }{x},\ Z_2=x\pde{\ }{x},\ Z_3=x^2\pde{\ }{x},\ 
Z_4=\pde{\ }{u},\ Z_5=u\pde{\ }{u},\ Z_6=u^2\pde{\ }{u}\},
\end{gather}
which span the 6-dimensional Lie algebra that
represents this projective transformation. 

Applying the condition
\begin{gather}
\label{cond-Inv-x}
Z_j^{(7)} \omega(x,u,u_x,u_{xx},u_{3x},u_{4x},u_{5x},u_{6x}, u_{7x})=0, \ j=1,2,\ldots,6,
\end{gather}
we obtain three invariants for the six-dimensional Lie algebra (\ref{Lie-Alg-6dim}), namely $\Omega_1$, 
$\Omega_2$ and $\Omega_3$ defined in Section 2 by (\ref{Omega-1}), (\ref{Omega-2}) and (\ref{Omega-3}), 
respectively, {\it viz.}
\begin{gather*}
%\label{Omega-1}
\Omega_1
=\frac{S_{xx}}{S^2}-\frac{5}{4}\frac{S_x^2}{S^3}+4\\[0.3cm]
%\label{Omega-2}
\Omega_2
=\frac{S_{3x}}{S^{5/2}}
-\frac{9}{2}\frac{S_xS_{xx}}{S^{7/2}}+\frac{15}{4}\frac{S_x^3}{S^{9/2}}\\[0.3cm]
%\label{Omega-3}
\Omega_3
=\frac{S_{4x}}{S^3}
-7\frac{S_xS_{3x}}{S^4}
+\frac{63}{4}\frac{S_x^2S_{xx}}{S^5}
+19\frac{S_{xx}}{S^2}
-\frac{315}{32}\frac{S_x^4}{S^6}
-\frac{95}{4}\frac{S_x^2}{S^3}
+34.
\end{gather*}
We remind that $S$ is the Schwarzian derivation (\ref{S-def}) and $S$ now depends only on $x$ as we are considering $u(x)$. Therefore the subscripts $x$ of $S$ denote ordinary derivatives here.

\smallskip

Using these invariants we can construct nonlinear ordinary differential equations (ODEs), all of which are invariant under the transformation (\ref{Mobius-u-x}). We restrict ourselves to ODEs that are algebraically solvable in terms 
of their highest derivatives, therefore we do not consider fully-nonlinear ODEs here. Under this restriction, we have the following three cases:

\strut\hfill

\noindent
{\bf Case 2.1:} {\it The fifth-order case}. There is only one invariant available here, namely $\Omega_1$, so we let
\begin{gather}
\Omega_1=\lambda
\end{gather}
which leads to the 5th-order equation
\begin{gather}
\label{2.1}
S_{xx}-\frac{5}{4}\frac{S_x^2}{S}+(4-\lambda)S^2=0,
\end{gather}
where $\lambda$ is any constant.

\strut\hfill

\noindent
{\bf Case 2.2:} {\it The sixth-order case}. Using the two invariants $\Omega_1$ and $\Omega_2$ we
consider
\begin{gather}
\Omega_2=F(\Omega_1),
\end{gather}
which leads to the 6th-order equation
\begin{gather}
\label{Eq-Case2.2}
S_{3x}-\frac{9}{2}\frac{S_xS_{xx}}{S}
+\frac{15}{4}\frac{S_x^3}{S^2}
-S^{5/2}\,F(\Omega_1)=0
\end{gather}
for any given function $F$ of its argument.

\strut\hfill

\noindent
{\bf Case 2.3:} {\it The seventh-order case}. Using the two invariants $\Omega_1$, $\Omega_2$ and $\Omega_3$ we consider
\begin{gather}
\Omega_3=F(\Omega_1,\Omega_2),
\end{gather}
which leads to the 7th-order nonlinear ODE
\begin{gather}
\label{Eq-Case-2.3}
S_{4x}
-7\frac{S_xS_{3x}}{S}
+\frac{63}{4}\frac{S_x^2S_{xx}}{S^2}
+19 SS_{xx}
-\frac{315}{32}\frac{S_x^4}{S^3}
-\frac{95}{4} S_x^2
+34S^3-S^3F(\Omega_1,\Omega_2)=0
\end{gather}
for any given function $F$ of its two arguments.

\strut\hfill

\noindent
{\bf Regarding Case 2.1:} We show that the 5th-order ODE in $u$, which is in fact the 2nd-order equation  (\ref{2.1})
in terms of $S$, namely
\begin{gather*}
%\label{ODE-5th-order-1}
S_{xx}-\frac{5}{4}\frac{S_x^2}{S}+(4-\lambda)S^2=0,
\end{gather*}
can easily be solved.
Note that in terms of $u$, equation (\ref{2.1}) takes the following form:
\begin{gather}
u_{5x}=\frac{1}{4u_x^3(2u_{3x}u_x-3u_{xx}^2)}
\left[
\vphantom{\frac{DA}{DB}}
10u_{4x}^2u_x^4
-40 u_{4x}u_{3x}u_{xx}u_x^3
+8\lambda u_{3x}^3u_x^3\right.\nn\\[0.3cm]
\label{5th-order-ODE-u}
\quad
-\left.
\vphantom{\frac{DA}{DB}}
9\left(\lambda-\frac{10}{3}\right) \left(
4u_{3x}^2u_{xx}^2u_x^2
-6u_{3x}u_{xx}^4u_x
+3 u_{xx}^6\right)
\right].
\end{gather}
Equation (\ref{2.1}) can be reduced to the 1st-order equation
\begin{gather}
\label{2.1-W}
S^2W \,W'-\frac{SW^2}{4}=(\lambda-4)S^2,
\end{gather}
where
\begin{gather}
\label{2.1-S}
W(S)=\frac{S_x}{S}.
\end{gather}
Here and below, primes denote ordinary derivatives with respect to $S$.
Recall that
\begin{gather}
\label{2.1-u}
S(x)=\frac{u_{3x}}{u_x}-\frac{3}{2}\frac{u_{xx}^2}{u_x^2},
\end{gather}
which can be solved for $u$ for any given $S(x)$. 
We conclude that the 5th-order equation (\ref{5th-order-ODE-u}) is in general solvable for any choice of the constant $\lambda$, leading to five arbitrary constant of integration: one arbitrary constant from the general solution of 
(\ref{2.1-W}), one additional arbitrary constant from the general solution of (\ref{2.1-S}), and three additional arbitrary constants from the general solution of (\ref{2.1-u}).

\strut\hfill

\noindent
{\bf Regarding Case 2.2:} The 6th-order ODE in $u$, which is in fact the 3rd-order equation (\ref{Eq-Case2.2})
in terms of $S$, namely 
\begin{gather*}
%\label{Eq-Case2.2}
S_{3x}-\frac{9}{2}\frac{S_xS_{xx}}{S}
+\frac{15}{4}\frac{S_x^3}{S^2}
-S^{5/2}\,F(\Omega_1)=0,
\end{gather*}
reduces to the 2nd-order equation 
\begin{gather}
\label{Eq-W-Case-2.2}
W''+\frac{\left(W'\right)^2}{W}-\frac{W'}{2S}
+\frac{W}{4S^2}=\frac{1}{S^{1/2}W^2}\,F(\Omega_1[W]),
\end{gather}
where 
\begin{gather}
W(S)=\frac{S_x}{S};\qquad \Omega_1[W]=WW'-\frac{W^2}{4S}+4.
\end{gather}
Furthermore, equation (\ref{Eq-W-Case-2.2}) can be written in the form
\begin{gather}
\label{Eq-V-Case-2.2}
S^2V''-\frac{SV'}{2}+\frac{V}{2}=2S^{3/2}V^{1/2}\,F(\Omega_1[V]),
\end{gather}
where 
\begin{gather}
V(S)=W^2(S)
\end{gather}
and
\begin{gather}
\Omega_1[V]=\frac{V'}{2}-\frac{V}{4S}+4.
\end{gather}
Note that (\ref{Eq-V-Case-2.2}) is the linear 2nd-order Euler equation in case $F=0$. We have not studied equation (\ref{Eq-V-Case-2.2}) for the case where $F\neq 0$ any further.

\strut\hfill

\noindent
{\bf Regarding Case 2.3:} The 7th-order ODE in $u$, which is in fact the 4th-order equation (\ref{Eq-Case-2.3})
in terms of $S$, namely 
\begin{gather*}
%\label{Eq-Case-2.3}
S_{4x}
-7\frac{S_xS_{3x}}{S}
+\frac{63}{4}\frac{S_x^2S_{xx}}{S^2}
+19 SS_{xx}
-\frac{315}{32}\frac{S_x^4}{S^3}
-\frac{95}{4} S_x^2
+34S^3-S^3F(\Omega_1,\Omega_2)=0
\end{gather*}
reduces to the following 3rd-order ODE:
\begin{gather}
S^3W^3W'''
+4S^3W^2W'W''
+S^3W(W')^3
+8S^2W^3W''
+11S^2W^2(W')^2
+11SW^3W'\nn\\[0.3cm]
-\frac{283}{32}W^4
-7W\left(
\vphantom{\frac{da}{db}}
S^2W^2W''+S^2W(W')^2+4SWW'+W^3\right)
+\frac{63}{4}
W^2\left(
\vphantom{\frac{da}{db}}
SWW'+W^2\right)\nn\\[0.3cm]
\label{Eq-Case-2-3-W}
+19S\left(
\vphantom{\frac{da}{db}}
SWW'+W^2\right)
-\frac{95}{4}SW^2
+34S^2
-S^2F(\Omega_1[W],\Omega_2[W])=0,
\end{gather}
where
\begin{gather}
W(S)=\frac{S_x}{S}
\end{gather}
and 
\begin{subequations}
\begin{gather}
\Omega_1[W]=WW'-\frac{W^2}{4S}+4\\
\Omega_2[W]=\frac{W}{S^{3/2}}
\left[
S^2WW''+S^2(W')^2-\left(\frac{9SW}{2}-4S\right)W'
+\frac{W^2}{4}\right].
\end{gather}
\end{subequations}
We have not studied equation (\ref{Eq-Case-2-3-W}) further.

\section{Concluding remarks}
In this paper we list all the evolution equations that are invariant under the given projective transformation (\ref{Mobius-u-x-t}) up to order seven. We use the invariants that we obtained for this transformation to construct the equations. The 3rd-order evolution equation in this class is symmetry-integrable and has the following hierarchy of symmetry-integrable equations:
%the hierarchy that originates from the 3rd-order fully-nonlinear equation 
%(\ref{Eq-Case1.1}), viz.
\begin{subequations}
\begin{gather}
\label{Hier-Memeber-1}
u_t=\lambda \frac{u_x}{S^{1/2}}\\[0.3cm]
\label{Hier-Memeber-2}
u_{t}=\lambda u_x\left[ \frac{1}{4}\frac{S_{xx}}{S^{5/2}}
-\frac{5}{16}\frac{S_x^2}{S^{7/2}}
+\left(\frac{1}{2}+k_1\right)\frac{1}{S^{1/2}}    \right]\\[0.3cm]
%\label{Hier-Memeber-3}
u_{t}=
\lambda u_x\left[
-\frac{1}{8}\frac{S_{4x}}{S^{7/2}}
+\frac{7}{8}\frac{S_xS_{3x}}{S^{9/2}}
+\frac{21}{32}\frac{S_{xx}^2}{S^{9/2}}
-\frac{231}{64}\frac{S_x^2S_{xx}}{S^{11/2}}
+\frac{1}{4}\left(2k_1+1\right)\frac{S_{xx}}{S^{5/2}}
\right.\nn\\[0.3cm]
\label{Hier-Memeber-3}
\quad
+\frac{1155}{64}\frac{S_x^4}{S^{13/2}}
-\frac{5}{16}\left(2k_1+1\right)\frac{S_x^2}{S^{7/2}}
\left.
+\frac{1}{4}\left(2k_1+1\right)^2\frac{1}{S^{1/2}}\right].
\end{gather}
\end{subequations}
The recursion operator for this hierarchy is given in Proposition 1. We show that this 3rd-order equation (\ref{Hier-Memeber-1})
can be mapped to the Schwarzian KdV by a hodograph-type transformation. We prove that there exist two fully-nonlinear 5th-order evolution equation that are both invariant under the transformation (\ref{Mobius-u-x-t}) and symmetry-integrable (see Proposition 2). We furthermore prove that there is no 6th-order evolution equation that is
symmetry-integrable (see Proposition 3). For the 7th-order case, we prove that the only quasilinear 7th-order evolution equation 
that is both invariant under (\ref{Mobius-u-x-t}) and symmetry-integrable with Lie-Bäcklund symmetry of order nine, 
is the equation (\ref{Hier-Memeber-3}) (see Proposition 4). We were not able to establish the symmetry-integrability, or its symmetry-nonintegrability, of the fully-nonlinear 7th-order equation (\ref{7th-order-FN-Q}), viz.
\begin{gather*}
%\label{7th-order-FN-Q}
u_t=\frac{u_x}{S^{1/2}}\frac{\Psi_{21}(\Omega_1,\Omega_2)}{\left[
\vphantom{\frac{da}{db}}
\Omega_3+\Psi_{22}(\Omega_1,\Omega_2)\right]^{3/4}}
+\Psi_{23}(\Omega_1,\Omega_2),
\end{gather*}
so the symmetry-integrability for this case is an open problem. 

\smallskip

By exploiting the optimal 1-dimensional subalgebras of the 7-dimensional Lie symmetry algebra spanned by (\ref{Lie-Alg-7dim}), we list the nontrivial symmetry reductions of the 3rd-order symmetry integrable equation (\ref{Hier-Memeber-1}). Remarkably, 
all ODEs so obtained are solvable and can hence be used to obtain explicit solutions for (\ref{Hier-Memeber-1}). The 5rh-order equation (\ref{Hier-Memeber-2}) and the 7th-order equation (\ref{Hier-Memeber-2}) can of course also be symmetry-reduced 
by the same 1-dimensional optimal subalgebras, but this is not done here.

\smallskip

We furthermore list all ODEs that are invariant under the projective transformation (\ref{Mobius-u-x-t}), where we make use of the invariants for this transformation to construct these equations.  In doing so, we restrict ourselves to those ODEs that can be solved in terms of their highest derivatives, therefore this class does not include fully-nonlinear equations. There exist three equations that belong to the mentioned class, namely the 5th-order equation in $u(x)$, which is conveniently written in terms of $S$ in (\ref{2.1}), viz.
\begin{gather*}
%\label{2.1}
S_{xx}-\frac{5}{4}\frac{S_x^2}{S}+(4-\lambda)S^2=0,
\end{gather*}
the 6th-order equation in $u(x)$, which is conveniently written in terms of $S$ in (\ref{Eq-Case2.2}), viz.
\begin{gather*}
%\label{Eq-Case2.2}
S_{3x}-\frac{9}{2}\frac{S_xS_{xx}}{S}
+\frac{15}{4}\frac{S_x^3}{S^2}
-S^{5/2}\,F(\Omega_1)=0,
\end{gather*}
and the 7th-order equation in $u(x)$, which is conveniently written in terms of $S$ in (\ref{Eq-Case-2.3}), viz.
\begin{gather*}
%\label{Eq-Case-2.3}
S_{4x}
-7\frac{S_xS_{3x}}{S}
+\frac{63}{4}\frac{S_x^2S_{xx}}{S^2}
+19 SS_{xx}
-\frac{315}{32}\frac{S_x^4}{S^3}
-\frac{95}{4} S_x^2
+34S^3-S^3F(\Omega_1,\Omega_2)=0.
\end{gather*}
We were able to solve the 5th-order ODE, reduce the 6th-order to the 2nd-order ODE (\ref{Eq-V-Case-2.2}), and reduce the 7th-order ODE to the 3rd-order ODE (\ref{Eq-Case-2-3-W}). These reductions are in part possible due to the fact that the three invariants from which they were constructed, namely (\ref{Omega-1}), (\ref{Omega-2}) and (\ref{Omega-3}), can be expressed in terms of the Schwarzian derivative $S$, which of course directly reduces the order of the equations by three. It is not clear to us whether the 2nd-order equation (\ref{Eq-V-Case-2.2}) or the 3rd-order equation (\ref{Eq-Case-2-3-W}) are exactly solvable or integrable equations, which might be the case for at least some choices of the functions $F$ that appear in both. It would be of interest to study these two equations in more detail.

Let us also point out two recent papers, namely \cite{E-E-79} and \cite{E-E-80},
where we have considered further projective transformations represented by the Lie algebras $sl(2,\mathbb R)$ and $sl(3,\mathbb R)$. In view of the current paper it could be of interest to revisit these transformations for a deeper study.

\section*{Acknowledgement} This work has been supported by GNFM of INdAM.

%\noindent
%{\bf Concluding remarks:} From our calculations we are tempted to conclude that the only evolution 
%equations up to order seven, 
%that are both invariant under (\ref{Mobius-u-x-t}) and symmetry-integrable are those that belong to the hierarchy of Proposition 1. Proving or disproving this is however still an open problem.
%, namely the equations of Cases 1, Case 2 and Case 4. 
%We aim to prove (or disprove) this statement in a future paper.

%%%%%%%%%%%%%%%%%%%%%%%
%%%%%%%%%%%%%%%%%%%%%%%%%%%
%%%%%%%%%%%%%%%%%%%%%%%%%%%%%%
%%%%%%%%%%%%%%%%%%%%%%%%%%%%%%%%%%

\label{lastpage}

\end{document}